\documentclass[10pt]{article}

\usepackage[utf8]{inputenc}
\usepackage[T1]{fontenc}
\usepackage{graphicx} % For figures
\usepackage{amsmath,amssymb} % Standard math
\usepackage{geometry}
\usepackage{setspace}
\usepackage{lineno} % Line numbers (optional but common in submission)
\usepackage{cite} % Numeric citations
\usepackage{hyperref} % For clickable links (safe for submission)
\usepackage{booktabs}
\usepackage[table]{xcolor}
\usepackage[dvipsnames]{xcolor}
\usepackage{xcolor}
\usepackage{svg}

\usepackage{xspace}
\usepackage{adjustbox}
\usepackage{textcomp}
\usepackage{ragged2e} % justify text for table footnote
\usepackage{tablefootnote}
\usepackage{threeparttable}
\usepackage{authblk}

\definecolor{lightblue}{rgb}{0.8,0.85,1}

\makeatletter
\@ifpackageloaded{xspace}{
  \xspaceaddexceptions{"'}
}{}
\makeatother
\newcommand*{\aefm}{AEFM\xspace}
\newcommand*{\swap}{Transition1x-2p3p4p\xspace}
\newcommand*{\tmc}{Transition1x-TMC\xspace}
\newcommand*{\xtb}{Transition1x-xTB\xspace}
\newcommand*{\ood}{Transition1x-OOD\xspace}

\title{Beyond the Training Domain:\\ Robust Generative Transition State Models for Unseen Chemistry}

\author[1,3,4]{Samir Darouich}
\author[1]{Jacob W. Toney}
\author[1,2]{Weiliang Luo}
\author[3]{Johannes Kästner}
\author[4]{Mathias Niepert}
\author[1,2]{Heather J. Kulik*}
\affil[1]{Department of Chemical Engineering, Massachusetts Institute of Technology, USA}
\affil[2]{Department of Chemistry, Massachusetts Institute of Technology, USA}
\affil[3]{Institute for Theoretical Chemistry, University of Stuttgart, Germany}
\affil[4]{Institute for Artificial Intelligence, University of Stuttgart, Germany}
\date{}

\begin{document}

\maketitle

\section{Abstract}

Transition states (TSs) govern the rates and outcomes of chemical reactions, making their accurate prediction a central challenge in computational chemistry. Although recent machine-learning models achieve near chemical accuracy in the prediction of TS structures and the associated reaction barriers for small organic reactions, their ability to generalize beyond the training domain remains largely unexplored. Here, we introduce targeted benchmarks to probe chemical and structural novelty in generative TS prediction. Building on Transition1x, a large-scale dataset of reactions involving small organic molecules, we construct curated extensions incorporating controlled elemental substitutions and diverse transition-metal complexes (TMC). These benchmarks reveal fundamental limitations of generative models in the generalization to previously unseen elements. As a result, they produce unphysical geometries and large energetic errors, even for reactions structurally similar to well-predicted organic systems. To address this challenge, we introduce a self-supervised pretraining strategy based on equilibrium conformers that exposes generative TS models to novel chemical environments prior to targeted fine-tuning. Across the newly proposed benchmarks, self-supervised pretraining substantially improves TS prediction for previously unseen systems, lowering the median root-mean-square-deviation (RMSD) of TS geometries on \tmc reactions from 0.39 to 0.19~\AA\ and reducing fine-tuning data requirements by up to 75\%, enabling reliable performance even in low-data regimes. Overall, the integration of generative TS models with self-supervised pseudo-reaction pretraining provides an efficient, scalable, and chemically robust framework for elucidating TSs well beyond the small organic molecule domain, establishing a foundation for investigating complex and catalytically relevant reaction landscapes with unprecedented breadth.

\section{Introduction}

Transition states (TSs) are the central determinants of chemical reactivity, governing both reaction rates and mechanistic pathways. A TS, formally defined as a first-order saddle point on the potential energy surface, requires sophisticated optimization procedures to locate, which, in the context of classical molecular modeling, generally fall into two classes. Single-ended methods\cite{banerjee_search_1985,baker_algorithm_1986,henkelman_dimer_1999} start from an initial guess and iteratively refine the geometry using gradient information, and in some cases Hessians. In contrast, double-ended approaches\cite{jonsson1998nudged,henkelman_climbing_2000,peters_growing_2004} seek the TS by constructing a reaction pathway between reactant and product structures and identifying the highest-energy point along this path. Although these strategies are widely employed, the underlying optimization algorithms are highly sensitive to the quality of the initial guess and often require many iterative steps to converge. In practice, poor initializations frequently lead to convergence failures. These limitations are further amplified by the high cost of electronic structure calculations, most prominently density functional theory (DFT)\cite{mardirossian2017thirty}, which are required to evaluate the potential energy surface and thus make large-scale reaction exploration computationally challenging.

Machine learning (ML) offers a compelling route to mitigate these computational bottlenecks. One line of work employs surrogate models to approximate the potential energy surface, enabling faster TS localization when integrated with conventional search algorithms. These surrogates include Gaussian process regressions\cite{pozun_optimizing_2012,koistinen_minimum_2016,koistinen_nudged_2017,denzel_gaussian_2018,denzel_gaussian_2019,garrido_torres_low-scaling_2019,heinen_transition_2022} as well as machine-learned interatomic potentials\cite{peterson_acceleration_2016,schreiner_neuralnebneural_2022,zhang2024exploring,wander_cattsunami_2024,yuan_analytical_2024,zhao_harnessing_2025, yin2025cats, tang2025accelerating}, both of which provide efficient, differentiable approximations to the underlying PES. While these models can dramatically reduce the cost of TS searches, their performance hinges on access to extensive non-equilibrium training data, especially in the vicinity of the TS itself, placing practical limits on how broadly they can be applied\cite{yuan_analytical_2024}.

In parallel to surrogate-assisted search strategies, a growing body of deep learning methods seeks to predict TS geometries directly\cite{pattanaik_generating_2020,jackson_tsnet_2021,zhang_deep_2021,choi_prediction_2023}. A major development in this direction is the emergence of generative models, which cast TS prediction as a conditional distribution-learning problem, where the goal is to sample plausible TS structures given the corresponding reactant and product configurations\cite{makos_generative_2021,duan_accurate_2023,kim_diffusion-based_2024,galustian_goflow_2025,duan_optimal_2025,hayashi_generative_2025, darouich2025adaptive}. OA-ReactDiff\cite{duan_accurate_2023} illustrates this approach by learning a joint distribution over reactant, TS, and product geometries with denoising diffusion\cite{ho2020denoising} and then reconstructing the TS through an inpainting step. Its successor, React-OT\cite{duan_optimal_2025}, builds on this framework by combining flow matching (FM)\cite{lipman_flow_2022,liu_flow_2022} and optimal transport, starting from an initial TS guess constructed via linear interpolation between reactant and product structures to improve sampling fidelity. Building on this line of work, Adaptive Equilibrium Flow Matching (\aefm)\cite{darouich2025adaptive} takes a complementary strategy by learning a contractive flow around the true TS manifold, which iteratively transports an initial TS guess toward a physically consistent geometry. Another class of methods bypasses the need for pre-aligned reactant–product pairs entirely by generating TS structures directly from 2D molecular graphs\cite{kim_diffusion-based_2024,galustian_goflow_2025,mark2025feynman}.

Although these generative methods have demonstrated impressive performance, their applicability remains constrained by the scope of available training data\cite{si2025transition}. Most current large-scale reaction datasets are dominated by organic chemistry and contain only a narrow set of light elements, primarily hydrogen, carbon, nitrogen, and oxygen\cite{schreiner_transition1x_2022,grambow2020reactants}. As a result, models trained on these benchmarks achieve near chemical accuracy within this limited domain but have yet to be demonstrated to generalize reliably to reactions involving diverse main-group elements or transition metals (TMs). These chemistries underpin many areas of real-world importance, including homogeneous and heterogeneous catalysis, energy conversion, and materials synthesis\cite{nandy_computational_2021,chen2022automated,tang2025accelerating,toney2025exploring}. The appeal of generative models is particularly strong for such complex systems, where the computational bottleneck often arises not just from complicated reaction pathways but also from the unfavorable scaling of DFT for large molecular sizes\cite{li2021computational}.

To assess the generalization of generative TS models we introduce a new benchmarking strategy that expands the existing Transition1x\cite{schreiner_transition1x_2022} dataset through atomic substitutions and the inclusion of transition metal complexes (TMCs). Transition1x comprises 10073 diverse organic reactions generated from an enumeration of 1154 reactants in the GDB7 dataset\cite{grambow2020reactants}, which contains small organic molecules with up to seven heavy atoms (C, N, and O) and a maximum of 23 atoms in total. Through systematic and yet chemically meaningful modifications of the molecular structures in Transition1x, we aim to probe model performance on previously unseen elements and reaction mechanisms. The resulting \swap dataset contains roughly 14000 reactions involving elements up to the third period while \tmc contains over 36000 reactions focused on ten catalytically relevant TMs, all computed at the GFN2-xTB\cite{Bannwarth2021} level of theory. Evaluations on these expanded benchmarks reveal significant limitations for generative models in out-of-distribution (OOD) chemistry as they often produce unphysical geometries with overlapping atoms and unreasonably short bonds, highlighting a fundamental failure mode for generative models\cite{peng_moldiff_2023,williams_physics-informed_2024, vost_improving_2025,wohlwend_boltz-1_2025, galustian_goflow_2025}.

To address these limitations, we introduce a novel self-supervised pretraining strategy for generative TS prediction that exploits conformers of equilibrium structures to generate pseudo-reactions. These equilibrium geometries are widely available across the periodic table and form the natural starting point for reaction exploration. The effectiveness of this strategy is evident from the model’s ability to extrapolate to new chemical compositions within in-distribution reaction types. For the \tmc benchmark, pretraining on conformers reduces the median RMSD between generated and ground truth TS structure dramatically, from 0.39~\AA\ for a vanilla model to 0.19~\AA\ for a model pretrained on \tmc conformers. Beyond improving performance on in-distribution reactions, this self-supervised pretraining strategy also greatly enhances generalization to OOD chemistry. It substantially reduces the amount of reaction-specific data required for fine-tuning, by up to 75\% as demonstrated on the \swap and \tmc benchmarks. The benefits are particularly evident in the ultra low-data regime, where combining only 50 targeted reactions with 100 pseudo-reactions allows React-OT to achieve a median RMSD of 0.15~\AA\ for a Pt-catalyzed methane activation, successfully capturing not only novel atom types but also an entirely new reaction mechanism. These results demonstrate that leveraging abundant equilibrium conformers for pretraining provides an effective, scalable, and generalizable strategy for reducing fine-tuning requirements while enabling reliable predictions on previously unseen systems.

Lastly, we demonstrate that generalization to novel systems with a DFT-level foundation, as in Transition1x, is feasible using GFN2-xTB as a compromise between accuracy and computational efficiency. By reoptimizing a subset of the generated data at the DFT level, we show that a \swap fine-tuned model achieves a median RMSD of 0.26~\AA\ relative to the corresponding DFT reference TSs. Leveraging the more than 700-fold speed-up of GFN2-xTB over DFT, this hybrid approach enables high-throughput reaction exploration while preserving near DFT level fidelity. To further close the gap to the desired accuracy, 1500 pseudo-reactions from the \tmc dataset are geometry-optimized at the DFT level and used for self-supervised pretraining before fine-tuning on \tmc. This procedure increases the structural similarity to DFT reference TSs by 10\%, reducing the RMSD from 0.47~\AA\ to 0.42~\AA, demonstrating an effective strategy to leverage DFT-quality conformers for generative TS prediction and reaction elucidation.

\section{Results}

\subsection{Overview}

\begin{figure}[ht]
\centering
\includegraphics[width=1.0\textwidth]{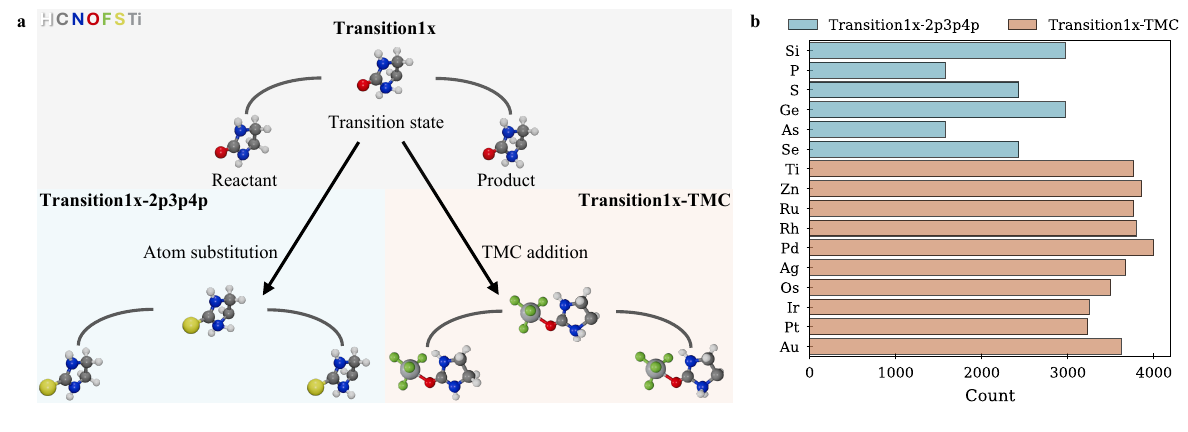}
\caption{
\textbf{Overview of data generation workflow for \swap and \tmc.}
\textbf{a} A Transition1x TS structure serves as the starting point for generating both benchmarks. In \swap, a single atom is replaced by an element from the same group up to the third period, producing systematically perturbed variants of the original reaction. In \tmc, the source TS is incorporated as an organic ligand into one of ten catalytically relevant TMCs. All modified geometries are refined to valid TS structures using the P-RFO algorithm, and corresponding reactants and products are obtained via IRC calculations. All structures are computed at the GFN2-xTB level of theory.
\textbf{b} Histogram of element frequencies, showing the number of reactions in which each element appears, colored according to the dataset (\swap or \tmc).
}\label{fig:dataset}
\end{figure}

To investigate whether React-OT and \aefm can transfer learned chemical knowledge beyond the small molecule domain, we evaluate their performance on the two newly introduced benchmarks \swap and \tmc. The \swap dataset introduces controlled perturbations to Transition1x reactions by randomly substituting a single heavy atom (C, N, or O) with another element from the same group but in a lower period, resulting in a set of 14,000 reactions that serves as a baseline for studying minor OOD variations and elemental extrapolation. Specifically, one-period-lower substitutions (e.g., C $\rightarrow$ Si) are classified as swap type 1, while two-period-lower substitutions (e.g., C $\rightarrow$ Ge) are classified as swap type 2. In contrast, \tmc represents a larger departure from the training domain, incorporating ten catalytically relevant TMs, namely: Ti, Zn, Ru, Rh, Pd, Ag, Os, Ir, Pt, and Au. For each metal, we selected the smallest synthetically accessible TMC, resulting in a dataset of over 36,000 reactions. This constitutes the first large-scale dataset for generative reaction modeling in organometallic chemistry. All structures are computed at the GFN2-xTB level of theory, providing an efficient balance between accuracy and computational cost. To establish a HCNO baseline at the GFN2-xTB level of theory, all Transition1x TS structures are re-optimized using the partitioned rational function optimization (P-RFO) method\cite{banerjee_search_1985,baker_algorithm_1986}, and intrinsic reaction coordinate (IRC)\cite{fukui1981path} calculations are subsequently performed to recover the corresponding reactant and product minima. Figure \ref{fig:dataset} provides an overview of the dataset creation (a) and the elemental composition of each dataset (b). To assess whether training on these more chemically diverse datasets improves extrapolation to previously unseen systems, we introduce three additional small benchmarks. The first, \ood, is derived from Transition1x by replacing hydrogen with fluorine or converting carbon–carbon double bonds into boron–nitrogen bonds. As a more stringent test, we further consider two catalytic reactions, a platinum-catalyzed methane activation and a rhodium-catalyzed ethylene activation. These systems introduce not only entirely new chemical elements, such as phosphorus and boron, but also reaction mechanisms that are absent from the training data. Further details can be found in Section~\ref{section:data_generation}.

\subsection{Analysis of limitations and failures}
\label{section:failure_analysis}

\begin{figure}[ht]
\centering
\includegraphics[width=0.9\textwidth]{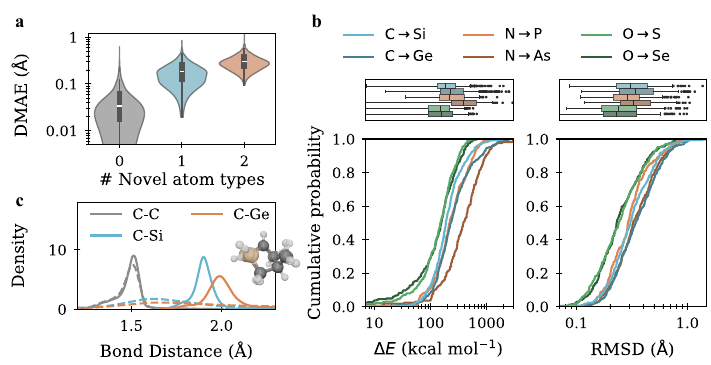}
\caption{
\textbf{Failure modes of the vanilla React-OT model on reactions involving novel chemistry.} \textbf{a} Mean bond distance mean absolute error (DMAE) of generated TS structures as a function of the number of previously unseen atom types, showing a rapid increase in error with growing elemental novelty. \swap introduces one novel atom type, while \tmc introduces up to two. \textbf{b} Cumulative probability distributions of structural and energetic errors for the \swap benchmark, illustrating the sharp performance degradation introduced by single-atom substitutions. \textbf{c} Ground-truth (solid) and predicted (dashed) bond-length distributions for representative C–C, C–Si, and C–Ge bonds, highlighting systematic geometric distortions in the model’s predictions.
}\label{fig:failure_analysis}
\end{figure}

Across the newly introduced \swap and \tmc benchmarks, both React-OT and \aefm exhibit significant limitations when applied to reactions outside their trained chemical domain. In particular, they are highly sensitive to the introduction of novel elements, even when the corresponding reactions remain closely aligned with their HCNO counterparts in the geometries of reactants, TSs, and products. This trend is reflected in Figure \ref{fig:failure_analysis}a, where the structural error of React-OT generated TS structures increases as additional novel atom types are introduced. While the vanilla React-OT model trained on \xtb performs reliably for HCNO reactions, even a single out-of-domain atom leads to a rapid deterioration in accuracy. Figure \ref{fig:failure_analysis}b further illustrates this behavior for the \swap benchmark, highlighting the sharp decline in performance once elemental substitutions extend beyond the HCNO space. Supplementary Figure~S4 demonstrates that \aefm follows the same trend, exhibiting a comparable degradation in performance. An overview of the performance of React-OT and \aefm is provided in Table~\ref{tab:performance} and Supplementary Table~S5.

To analyze the influence of chemical identity, we specifically selected reactions in which single-atom substitutions or the addition of a TMC produce only minimal structural changes, keeping the underlying reaction mechanism effectively unchanged. A more detailed definition for similar reactions is provided in Supporting Section~6.3. Within the \swap benchmark, this procedure yielded 40 reactions that remain structurally aligned with entries in the \xtb dataset. Even the introduction of a single novel element substantially degrades React-OT’s accuracy, raising the median RMSD from 0.04~\AA\ for the corresponding HCNO reactions to 0.18~\AA. The \tmc benchmark follows the same trend, where structurally analogous reactions containing TMCs lead to poor performance and an increased median RMSD of 0.39~\AA. In contrast, the corresponding HCNO analogues are again predicted with high accuracy, with a median RMSD of 0.05~\AA\ (see Supplementary Table~S3). 

Many TS candidates appear plausible based on global metrics such as RMSD, however even small deviations in specific bond lengths can incur substantial energetic penalties. Figure~\ref{fig:failure_analysis}c illustrates this behavior, where the predicted C–C bond length distribution closely matches the reference data, while predictions for C–Si and C–Ge bonds show substantial deviations. Instead of capturing their true peaks at 1.90~\AA\ and 2.01~\AA, React-OT generates a broadened distribution with only a weak peak near 1.6~\AA, much closer to the organic C–C peak at 1.49~\AA\ than to the correct bond lengths. Global structural metrics for \aefm are comparable to those of React-OT, while bonds involving unseen elements exhibit similarly broadened and systematically shifted distributions. 

These findings indicate that training exclusively on an HCNO dataset such as Transition1x is insufficient to enable robust generalization to reactions involving novel elements, even when the underlying reaction mechanism remains unchanged. To assess whether exposure to chemically more diverse data improves extrapolation, both React-OT and \aefm were fine-tuned on the \swap and \tmc datasets, respectively. Dataset splits were constructed by inheriting the original Transition1x train–test partition, such that any modified reaction in \swap or \tmc follows the split assignment of its parent reaction. This procedure results in 18594 and 32633 training reactions for \swap and \tmc, respectively. Details of the fine-tuning protocol are provided in Supplementary Section~S1.2. To assess whether chemically more diverse training data improves extrapolation, we evaluated the fine-tuned models on additional benchmarks. On \ood, the vanilla React-OT model achieves a median RMSD of 0.15~\AA, while the \swap-fine-tuned model reaches 0.19~\AA. Both models exhibit the same behavior observed earlier, predicting bond lengths for novel atom types based on biases from related interactions in the training set. A detailed analysis of a representative boron–hydrogen bond is provided in Supplementary Section~S3.2. A similar trend is observed for catalytic benchmarks. The vanilla React-OT model attains median RMSDs of 0.33~\AA\ for the Pt-catalyzed system and 0.24~\AA\ for the Rh-catalyzed system, while the \tmc pretrained model shows limited transferability, with median RMSDs of 0.30~\AA\ and 0.32~\AA, respectively (Supporting Table~S9). Focusing on the reactive center, consisting of Rh, the coordinated ethylene, and adsorbed H$_2$, the \tmc pretrained model improves slightly, reaching a median RMSD of 0.17~\AA, compared to 0.20~\AA\ for the vanilla model. This indicates that pretraining on Rh-containing TMCs enhances the description of the metal center and its immediate coordination environment. However, generalization to chemically novel ligands remains limited, as reflected by persistent errors for phosphorus-containing species. Overall, these results suggest that broader training coverage alone does not ensure reliable extrapolation to novel chemical systems, highlighting the need for approaches that more directly promote transferability in generative reaction models.

\subsection{Conformer-based pretraining enables data-efficient generalization}

\begin{figure}[ht]
\centering
\includegraphics[width=0.9\textwidth]{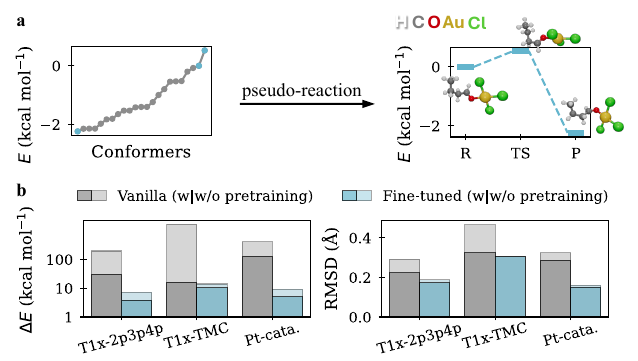}
\caption{
\textbf{Conformer-based pseudo-reaction pretraining.} 
\textbf{a} Construction of pseudo-reactions from sets of conformers. The conformer with the second-highest energy is assigned as the reactant, the highest-energy conformer as the TS, and the lowest-energy conformer as the product.
\textbf{b} Effect of self-supervised pretraining using 2500 (\swap and \tmc) and 100 (Pt catalyzed) pseudo-reactions constructed from equilibrium minima of the corresponding datasets on the median performance of vanilla and fine-tuned React-OT models. The fine-tuned results are obtained using reduced training sets comprising 10\% of \swap and \tmc and 50 reactions for the Pt-catalyzed benchmark. Transparent bars indicate models trained without pretraining, while solid bars denote models incorporating self-supervised pretraining.
}\label{fig:crest}
\end{figure}

To address the limited generalization of generative reaction models to novel elements, we introduce conformer-based pretraining as a data-efficient strategy to promote transferability beyond the training distribution. Unlike TSs, which are scarce and computationally expensive to obtain, equilibrium conformers are abundant across chemical space and can be generated efficiently through standard geometry optimizations. Leveraging this abundance, we formulate a conformer-based self-supervised learning scheme that constructs pseudo-reactions from equilibrium structures, exposing the model to diverse chemical environments prior to supervised fine-tuning (see Section~\ref{section:crest_pretraining}). For each molecule, conformers are ranked by their relative energies and three representative structures are selected to mimic a reaction trajectory. The highest-energy conformer is treated as a pseudo-TS, the second-highest as the pseudo-reactant, and the lowest-energy conformer as the pseudo-product, as illustrated in Figure~\ref{fig:crest}a. These pseudo-reactions do not correspond to actual chemical transformations but capture realistic geometrical variations and energetic progression along a reaction-like coordinate. When used to pretrain OAReactDiff, these pseudo-reactions expose the model to new chemical environments and promote the learning of chemically realistic structural patterns. Using OAReactDiff in this self-supervised pretraining setup is particularly natural because it models the joint distribution of the reaction, allowing the model to learn not only from pseudo–TSs but also from the corresponding pseudo–reactant and pseudo–product structures. The pretraining protocol is implemented in a two-stage procedure, starting from a pretrained OAReactDiff checkpoint. In the baseline case, the OAReactDiff model is directly fine-tuned on the target reaction data using the React-OT FM loss. In the self-supervised pretraining workflow, the checkpoint is first trained on pseudo-reactions derived from equilibrium conformers of the target chemical domain, and subsequently fine-tuned on the corresponding real reactions using the React-OT FM loss.

The advantages of pseudo-reaction pretraining become clear when comparing the vanilla React-OT model, trained only on \xtb HCNO chemistry, with models that are first pretrained on pseudo-reactions derived from \swap or \tmc equilibrium structures and then fine-tuned on the \xtb reactions. In addition to structural accuracy measured by RMSD, we assess energetic fidelity by computing the single-point energy difference between the ground truth TS and the generated structure using GFN2-xTB. The vanilla React-OT model produces highly inaccurate structures for \swap, with a median RMSD of 0.29~\AA\ and a median energetic error of 196.99~kcal~mol$^{-1}$. Pretraining with 500 pseudo-reactions lowers the median RMSD to 0.23~\AA, and expanding the pretraining set to 2500 pseudo-reactions reduces it to 0.22~\AA, accompanied with a median energetic error of 30.99~kcal~mol$^{-1}$. The impact of the self-supervised pretraining is even more pronounced for \tmc, as shown in Figure~\ref{fig:crest}b. The vanilla React-OT model struggles severely with these systems, producing highly unrealistic structures with a median RMSD of 0.47~\AA. By comparison, pretraining with 500 pseudo-reactions reduces the median RMSD to 0.34~\AA, and 2500 pseudo-reactions further lower it to 0.32~\AA\ with a median energetic error of 16.24~kcal~mol$^{-1}$. To understand the origin of these improvements, we focus once more on reactions that remain structurally similar to the original \xtb set but contain elements not present during training. The vanilla React-OT model struggles with such unfamiliar elements and frequently produced unrealistic geometries, despite being able to predict the same reactions correctly when restricted to its HCNO domain. The self-supervised pseudo-reaction pretraining mitigates this limitation by exposing the model to chemically plausible structural patterns for these novel elements. Once relieved from the challenge of extrapolating to entirely unfamiliar chemistry, the model can concentrate on the underlying mechanistic prediction task and make effective use of the knowledge it has already acquired. As a result, the pretrained model produces TS structures that lie substantially closer to the ground truth. On the \swap benchmark, the median RMSD is reduced from 0.18~\AA\ for the vanilla model to 0.10~\AA\ after pretraining on 2500 pseudo-reactions, corresponding to a 42\% improvement. Consistent with this overall structural improvement, the median bond distance mean absolute error (DMAE) decreases from 0.09~\AA\ to 0.06~\AA, indicating a more accurate representation of bonded distance distributions. On \tmc, the median RMSD decreases from 0.39~\AA\ to 0.19~\AA\, yielding a 51\% improvement, while the median DMAE reduces from 0.26~\AA\ to 0.10~\AA\ (Supporting Table~S3).

Beyond enhancing OOD performance, self-supervised pseudo-reaction pretraining substantially enhances data efficiency. For \swap, a pretrained model trained on only 25\% of the dataset performs comparably to a model trained on the full dataset. Similarly, for \tmc, pretraining allows the model to achieve the same accuracy as a fully trained model using only 50\% of the data (Supplementary Figure~S6). This efficiency is particularly valuable for elucidating novel reaction mechanisms, where high-quality reaction data is limited. Even a modest pretraining set of 100 pseudo reactions supplemented with only a small number of real reactions, can bring the model’s performance close to the desired accuracy. For instance, in the Rh-catalyzed ethylene hydrogenation, a React-OT model fine-tuned on 50 reactions yields a median RMSD of 0.04~\AA\ with a median energetic error of 3.65~kcal~mol$^{-1}$. Pretraining with just 100 pseudo-reactions and fine-tuning on the same 50 reactions reduces this error to 0.03~\AA\ and 2.45~kcal~mol$^{-1}$. A similar trend is observed for the Pt-catalyzed methane activation. The model fine-tuned with 50 reactions attains a median RMSD of 0.16~\AA\ together with a median energetic error of 9.04~kcal~mol$^{-1}$. Incorporating 100 pseudo-reactions reduces this to 0.15~\AA\ and 5.06~kcal~mol$^{-1}$, as shown in Figure~\ref{fig:crest}b and Supplementary Table~S9.

\begin{table*}[ht]
\centering
\caption{Structural and energetic errors for variants of React-OT.}
\vspace{0.5em}
\label{tab:performance}
\begin{adjustbox}{max width=\linewidth}
\begin{tabular}{@{\extracolsep\fill}llllllll}
\toprule
Approach & Dataset & \multicolumn{2}{c}{RMSD (\AA)} & \multicolumn{2}{c}{DMAE (\AA)} & \multicolumn{2}{c}{$|\Delta E_{\text{TS}}|$ (kcal mol$^{-1}$)} \\
\cmidrule(lr){3-4} \cmidrule(lr){5-6} \cmidrule(lr){7-8}
& & Mean & Median & Mean & Median & Mean & Median \\
\midrule
Vanilla React-OT & \swap & 0.33  & 0.29 & 0.18 & 0.17 & 282.91 & 196.99 \\
Vanilla React-OT\textsuperscript{\emph{a}} & \swap & 0.28 & 0.22 & 0.14 & 0.12 & 47.23  & 30.99  \\
Vanilla React-OT & \tmc & 0.51 & 0.47 & 0.31 & 0.29 & 2591.42 & 1693.94 \\
Vanilla React-OT\textsuperscript{\emph{a}} & \tmc & 0.38 & 0.32 & 0.18 & 0.15 & 28.32 & 16.24 \\
Fine-tuned React-OT & \swap & 0.20 & 0.14 & 0.09 & 0.06 & 6.68  & 2.97 \\
Fine-tuned React-OT\textsuperscript{\emph{a}} & \swap & 0.20 & 0.14 & 0.09 & 0.06 & 5.96  & 2.72 \\
% Fine-tuned React-OT + \aefm & \swap & 0.20  & 0.12 & 4.20  & 0.62 \\
Fine-tuned React-OT & \tmc & 0.30 & 0.24 & 0.13 & 0.10 & 10.45 & 6.68 \\
Fine-tuned React-OT\textsuperscript{\emph{a}} & \tmc & 0.30 & 0.24 & 0.13 & 0.10 & 9.80 & 6.25 \\
% Fine-tuned React-OT + \aefm & \tmc & 0.26 & 0.15 & 4.16  & 0.72 \\
\bottomrule
\end{tabular}
\end{adjustbox}
\footnotesize
\justify
\textsuperscript{\emph{a}} self-supervised pretraining with 2500 pseudo-reactions of the corresponding dataset. \\
\end{table*}

\subsection{Transferability of semi-empirical methods}

Up to this point, all analyses are performed at the GFN2-xTB level of theory. We therefore ask whether a model pretrained on high-quality DFT data can be adapted to new chemical environments using only GFN2-xTB structures, while still maintaining DFT-level accuracy. Successfully doing so would allow rapid large-scale screening and subsequent TS refinement without compromising the fidelity of DFT calculations.

\begin{figure}[ht]
\centering
\includegraphics[width=0.9\textwidth]{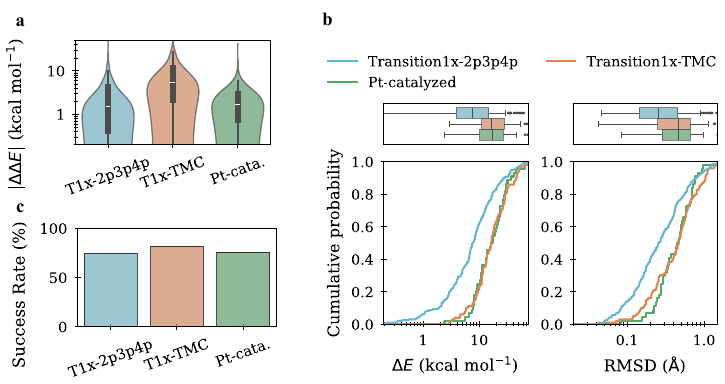}
\caption{
\textbf{Validation at the DFT level of theory.} 
\textbf{a} Absolute difference between energetic error evaluated at GFN2-xTB and DFT level of theory. Errors increase at the DFT level for all datasets, with the effect more pronounced for \tmc.
\textbf{b} Cumulative probability distributions of structural and energetic errors between the generated samples and their corresponding reference DFT TS.
\textbf{c} Success rate of converging to the desired reference DFT TS structure after DFT optimization of the generated samples. A successful convergence is defined as achieving an RMSD of 0.05~\AA\ or lower.
}\label{fig:transferability_gfn}
\end{figure}

To address this question, we evaluate the consistency between energetic differences computed at the GFN2-xTB and DFT levels. Therefore, Single-point DFT energies are computed for all TS structures generated at the GFN2-xTB level as well as for structures predicted by React-OT. The energetic differences, $\Delta E_\text{TS}$, are compared across the two levels of theory. Across all datasets, including \swap, \tmc, Rh catalysis, and Pt catalysis, the mean deviation between $\Delta E_\text{TS}^\text{DFT}$ and $\Delta E_\text{TS}^\text{GFN2-xTB}$ remains below 25\%. Dataset-specific deviations are reported in Supplementary Table~S10, and the distribution of absolute energetic differences is shown in Figure~\ref{fig:transferability_gfn}a. This agreement indicates that GFN2-xTB capture a substantial portion of the underlying DFT potential energy surface and can serve as a reliable and computationally efficient foundation for training generative reaction models intended for large-scale screening. 

To further assess the structural fidelity of React-OT predictions, we reoptimize a subset of TS structures from the \swap, \tmc, and Pt catalyzed datasets, originally generated at the GFN2-xTB level, at the DFT level using the P-RFO algorithm. For the \swap dataset, 200 TS structures were selected, comprising 100 examples from each swap class, of which 194 successfully converged to DFT TSs. From the \tmc dataset, 100 TS structures were chosen for refinement, including 10 for each TM, with 99 converging under the same protocol. For the Pt-catalyzed system, 50 TS structures were refined, 44 of which converged successfully. For the \swap dataset, the resulting comparison shows that React-OT predictions align closely with their DFT-optimized counterparts, with a median RMSD of 0.26~\AA\ and a median energy difference of 7.67~kcal~mol$^{-1}$. The \tmc dataset presents a more diverse chemical landscape, however the model maintains reasonable agreement, yielding a median RMSD of 0.47~\AA\ and a median energetic deviation of 16.34~kcal~mol$^{-1}$. A similar behavior is observed for the Pt catalyzed system, with a median RMSD of 0.46~\AA\ and a median energetic difference of 16.99~kcal~mol${-1}$ (Figure~\ref{fig:transferability_gfn}b and Supplementary Table~S12). In a second analysis, the React-OT predicted TSs are refined at the DFT level to examine whether they reliably converge toward the reference DFT TS structures. For the \swap dataset, 188 of the 200 predicted TSs converge successfully, with 75\% reaching the corresponding DFT-optimized structures. Structures are considered identical if the RMSD is below 0.05~\AA. In the \tmc dataset, 93 React-OT predictions converge, and 82\% match the reference TS geometries under the same criterion. For the Pt-catalyzed benchmark, 37 out of 50 selected samples converged with a matching rate of 76\% (Figure~\ref{fig:transferability_gfn}c and Supplementary Table~S13).

To elevate the DFT level accuracy of the model, we reoptimized equilibrium conformers generated from \tmc dataset at the $\omega$B97X/def2-SVP level of theory. Since these structures correspond to equilibrium geometries rather than complete reaction paths, a large number can be processed at relatively low computational cost. In total, 1500 pseudo-reactions, corresponding to 4500 structures, were successfully optimized. Pretraining on these 1500 DFT level pseudo-reactions, followed by fine-tuning on the \tmc dataset, reduces the median RMSD from 0.47 to 0.42~\AA\ and decreases the median energetic deviation from 16.34 to 15.48~kcal~mol$^{-1}$ relative to a model trained solely on \tmc (Supplementary Table~S12). This indicates that leveraging DFT-optimized conformers can effectively strengthen pretraining, offering a valuable strategy for improving model accuracy.

\section{Discussion}

This work systematically exposes the limitations of generative models for TS prediction when applied to chemical spaces beyond their training domain. To assess this, we developed a benchmark strategy probing both elemental novelty and previously unseen reaction mechanisms by swapping atom types in Transition1x TS structures and incorporating TMCs. We find that the presence of unfamiliar elements is a primary factor limiting the model’s ability to leverage learned mechanistic representations, leading to reduced structural and energetic accuracy. To address this, we introduced a pseudo-reaction self-supervised pretraining strategy based on equilibrium structures. These geometries are abundant, readily accessible, and naturally serve as starting points for reaction elucidation, providing an effective way to expose the model to new chemistries while substantially reducing the amount of reaction data required for fine-tuning. When supplemented with even a small reaction dataset, this pretraining enables accurate TS predictions for complex catalytic systems, demonstrating adaptability to previously inaccessible chemical regimes. Despite these advances, the scarcity of high-quality reaction data remains a central challenge. Semi-empirical methods offer an effective route to extend generative models to new chemical environments while maintaining meaningful accuracy at the DFT level. Incorporating modest amounts of DFT-level information further improves predictive fidelity, as pseudo-reactions optimized at this level consistently enhance performance. Together, these findings point to a hybrid data paradigm in which broad semi-empirical coverage is reinforced by targeted high-accuracy calculations, providing a practical path toward reliable TS prediction across chemically diverse systems. Future developments could exploit generative models capable of producing DFT-level conformers, which can be seamlessly integrated with self-supervised pretraining to accelerate exploration. In this context, uncertainty-aware active learning could guide the selection of the most informative reactions, complementing pseudo-reaction pretraining to further maximize data efficiency. Collectively, these strategies have the potential to reduce reliance on costly quantum-chemical calculations and facilitate the efficient elucidation of previously inaccessible reactions. 

\section{Method}

\subsection{Data generation}
\label{section:data_generation}

\subsubsection{\swap}

As a first test for novel chemistry, permutations of atom types within the same group are performed for Transition1x\cite{schreiner_transition1x_2022} reactions, where in each reaction only a single atom is substituted at a time. Two types of substitutions are considered. In the first, a carbon is replaced by silicon, nitrogen by phosphorus, and oxygen by sulfur. In the second, carbon is replaced by germanium, nitrogen by arsenic, and oxygen by selenium. As these substitutions remain within the same periodic group, broadly similar chemical behavior is expected. For each TS in Transition1x, a random permutation source type and afterwards a random atom of the corresponding type is chosen. Hence, first the source type, consisting of either carbon, nitrogen, or oxygen, and afterwards an atom of this type in the molecule is chosen. The same atom is chosen for swap 1 as well as for swap 2. The modified TS is reoptimized using the partitioned rational function optimization (P-RFO) algorithm\cite{banerjee_search_1985,baker_algorithm_1986}, and the associated minima are located through intrinsic reaction coordinate (IRC)\cite{fukui1981path} calculations. In addition, all original DFT Transition1x reactions are recomputed at the GFN2-xTB\cite{Bannwarth2021} level using the same protocol. A detailed description of the procedure is provided in Section~\ref{section:opt_procedure}. To ensure a consistent basis for comparison across novel atom types, only reactions for which the corresponding source reaction converged successfully under both swap types are retained. If a reaction converged for only one of the two substitutions, both resulting swapped reactions are excluded from the dataset. This filtering ensures that each retained pair of swapped reactions (types 1 and 2) originates from the same source TS and allows a controlled analysis of the impact of single-atom substitutions. This procedure yields a total of 20922 reactions, comprising 6974 original HCNO reactions and 6974 reactions for each swap type. During training, the corresponding \xtb source reactions are included to augment the dataset, whereas evaluation is performed exclusively on the substituted reactions. Additional details regarding the inclusion of \xtb reactions in the \swap dataset are provided in Supporting Section~S1.2.

\subsubsection{\tmc}

To analyze more complex chemistry, small transition metal complexes (TMCs) are created using the TS structures from the Transition1x dataset. Therefore, ten TMs from the 3d, 4d, and 5d series of the periodic table (i.e., Ti, Zn, Ru, Rh, Pd, Ag, Os, Ir, Pt, and Au) are chosen, which are both catalytically relevant and likely to exhibit a low-spin ground state. To make sure the TMCs are synthetically-accessible, ten crystal structures from the Cambridge Structural Database(CSD) \cite{groom2016CSD} which are small, representative of each metal, and have a known charge and metal oxidation state are selected. These structures tend to be homoleptic metal chloride complexes, with the exception of dicyano-silver, dicarbonyl-(dichloro)-rhodium, and hexafluoro-titanium. For each of these ten TMCs, anionic ligands of known charge (i.e., either Cl$^-$, F$^-$, or CN$^-$) are removed until the complex reached an overall net neutral charge. Leading to the following neutral TMCs: Ti(IV)F$_4$, Zn(II)Cl$_2$, Ru(III)Cl$_3$, Rh(I)Cl(CO$_2$)$_2$, Pd(II)Cl$_2$, Ag(I)CN, Os(IV)Cl$_4$, Ir(IV)Cl$_4$, Pt(IV)Cl$_4$, Au(IV)Cl$_3$. Using these TMCs as scaffolds, the 7516 unimolecular TS from Transition1x are treated as organic ligands which may coordinate to a TM with available binding sites. Coordinating atoms for each of the 7453 organic ligands with parsable SMILES strings\cite{weininger1988SMILES} are predicted using the graph neural network model pydentate\cite{toney2025pydentate1,toney2025pydentate2}. Only the 4573 ligands predicted to exhibit monodentate coordination to the metal are retained. These metal–ligand complexes are prepared by aligning the vector formed by the predicted coordinating atom and center of mass with the vector formed by the TM and the average coordinates of its deleted anionic ligands. The resulting new metal–ligand bond length is set to a distance equal to the sum of the covalent radii between the metal and coordinating atom. All structure generation is performed with molSimplify\cite{molSimplify,terrones2025molsimplify}. After adding the TS structure to the TMC as an additional organic ligand, the reaction protocol outlined in Section~\ref{section:opt_procedure} is used to find new TS structures and their corresponding minima. In order to have the most diverse dataset, every converged reaction is kept, leading to 36446 unique organometallic reactions.

\subsubsection{Additional Benchmark Sets}

To further assess the transferability of models trained on the \swap and \tmc datasets, we construct three additional benchmark sets that introduce distinct forms of chemical novelty. The first, termed \ood, is derived from Transition1x by introducing elemental and bonding modifications beyond the original training domain. Specifically, hydrogen atoms are substituted by fluorine, and carbon–carbon double bonds within six-membered rings are replaced by boron–nitrogen bonds. This procedure yields 97 hydrogen-to-fluorine substitutions and 25 carbon–carbon to boron–nitrogen substitutions, resulting in a total of 122 reactions. As a demonstration in a practical catalytic context, a dataset of 2073 rhodium-catalyzed ethylene hydrogenation reactions\cite{tang2025accelerating} is used. This dataset includes 3D TS guesses generated by GENiniTS-RS. Applying the reaction optimization procedure to these initial TS structures yields 1839 successfully converged reactions. Redundant reactions are removed by filtering for similarity, defined as reactions for which the RMSD between reactants, products, and TSs is below 0.1~\AA, resulting in 1619 unique reactions. The dataset comprises a diverse set of elements, including Rh, Cl, Br, P, and F. As second application, a dataset consisting of platinum-catalyzed methane activation reactions\cite{chen2022automated}, providing 317 DFT TS structures is examined. The TS structures are optimized at the GFN2-xTB level using the reaction procedure, leading to 195 converged reactions. This dataset also contains a variety of elements, including Pt, I, Cl, Br, P, and F.

\subsubsection{Optimization procedure}
\label{section:opt_procedure}

GFN2-xTB is employed for all generated structures, providing an efficient balance between accuracy and computational cost. TS structures are located using the P-RFO algorithm, accompanied by an IRC calculation to determine the corresponding minima. This step is essential, as starting from a DFT TS structure does not guarantee that the resulting xTB TS connects to the same reactant and product minima. ASE\cite{ase-paper} along with the Sella package\cite{hermes2022} and the tblite interface is utilized for TS optimization and IRC calculations. The TS optimization is considered converged when the maximum force falls below 0.001 eV$\text{\AA}^{-1}$ within a maximum of 500 iterations. For IRC calculations, a convergence threshold of 0.1 eV$\text{\AA}^{-1}$ is applied, also with a maximum of 500 iterations. The minima obtained from the IRC are subsequently relaxed using gradient descent with a stricter threshold of 0.05 eV$\text{\AA}^{-1}$ and 500 iterations. A reaction is classified as converged when all TS and minima criteria are satisfied and the IRC minima are distinct, defined as an RMSD exceeding 0.1~\AA\ between the located minima. For all Transition1x derivatives, the data is partitioned following the original split established by OAReactDiff. Accordingly, any reaction that is modified, either through atom swaps or by incorporating a TMC, retains the split assignment of its corresponding parent reaction. Under this scheme, the \swap dataset comprises 18594 training reactions, which include both the source \xtb reactions and the substituted reactions, and 1552 test reactions drawn exclusively from swap type~1 and type~2 substitutions. The \tmc dataset contains 32633 training and 3813 test reactions. For \aefm, 10\% of the training data is further held out for early stopping.

\subsubsection{Quantum chemical validation}

For DFT TS optimization, the Sella package with the ASE ORCA 6.0.1\cite{ORCA} calculator is used, with varying level of theory. All reactions derived from the Transition1x dataset are performed with the $\omega$B97x/def2-SVP\cite{chai2008systematic,weigend2005balanced} level of theory. The bigger basis set compared to the original Transition1x calculations is needed to account for the newly introduced atomic types. For the catalytic datasets, the same level of theory is used as in the respective original publication, $\omega$B97x-D3/def2-SVP\cite{chai2008systematic,grimme2010consistent,weigend2005balanced}, and PBE0 D3BJ/def2-SVP\cite{adamo1999toward,becke2005density,weigend2005balanced} for rhodium catalysis, and platinum catalysis, respectively. For the \swap dataset, 100 reactions are selected for DFT optimization. Performing this procedure for both swap types resulted in 194 out of 200 TS structures successfully converging. In the \tmc dataset, 10 reactions that converged for each TM are chosen, yielding 100 DFT-optimized TSs, of which 99 converged. For the Pt-catalysis benchmark, 50 GFN2-xTB TS structures are reoptimized at the DFT level, as the xTB mechanisms do not necessarily reproduce the original DFT reaction pathways, and 44 TSs converged. For the Rh-catalyzed reactions, the system size (median 44 atoms) rendered full TS DFT optimization impractical. As a result, only single-point DFT energies are computed.

\subsection{Pseudo-reactions for self-supervised pretraining}
\label{section:crest_pretraining}

Large-scale reaction datasets remain scarce for chemically diverse systems, posing a major bottleneck for machine learning–based reaction elucidation. In contrast, equilibrium structures are readily available across a broad range of chemistries. Leveraging this disparity, we introduce a self-supervised pretraining strategy that constructs pseudo-reactions from conformers of equilibrium structures. As a starting point, we generate conformers of the equilibrium structure and convert them into pseudo-reactions by assigning the second-highest-energy conformer as the reactant, the highest-energy conformer as the TS, and the lowest-energy conformer as the product. Conformers are obtained using the Conformer–Rotamer Ensemble Sampling Tool (CREST)\cite{pracht2020automated,pracht2024crest}, with the energy window for conformer generation expanded to 20~kcal~mol$^{-1}$ to ensure broad coverage of conformational space. While CREST is used throughout this work, any conformer sampling approach, including machine-learned conformer generators, could be employed within this framework. Conformer searches are performed on a subsample of minima for each dataset. For \swap and \tmc, up to 2500 pseudo-reactions are generated at the GFN2-xTB level of theory, whereas for the OOD datasets, 100 pseudo-reactions are generated at the GFN-FF\cite{spicher2020robust} level. Both reactant and product minima are considered for the conformer search. The structural similarity between each pseudo-reaction conformer and the corresponding structure in the original reaction is evaluated. Supporting Table~6-8 reports RMSD statistics for all structure combinations, indicating RMSD values above 1~\AA, confirming that pseudo-reactions are structurally distinct from the original reactions. The pseudo-reactions are subsequently used with OAReactDiff to pretrain a model on novel chemistries and structures. This pretrained model is then fine-tuned on the available data using the React-OT task.

\subsection{Generative models}

OAReactDiff\cite{duan_accurate_2023} is an object-aware diffusion model, learning the joint probability distribution of the entire reaction, namely the reactant, TS, and product structure. It can be used for double-ended TS search by using inpainting given the reactant and product. React-OT\cite{duan_optimal_2025} is the current state-of-the-art generative model for double ended TS generation. It leverages object-aware flow matching (FM) in combination with an optimal transport path. As source distribution a linear combination of reactant and product structure is used. Adaptive Equilibrium Flow Matching (AEFM)\cite{darouich2025adaptive} uses a novel version of FM, called equilibrium FM, learning a flow field which is iteratively applied until a fixed-point convergence is achieved. AEFM is not conditioned on reactant or product structures, and serves as an analogy of surface walking methods to find TS structures given an initial guess of the TS. In this work, the linear interpolation of reactant and product is used as initial guess. To train AEFM, additional 10 percent of the training data are reserved for early stopping. More training details for all models are reported in the Supporting Section~S1.1.

\subsection{Metrics}

The RMSD between two molecular structures is determined by first aligning the molecules $\mathbf{x}_1$ and $\mathbf{x}_2$ using the Kabsch algorithm and then computing:
\begin{equation}\label{eq:rmsd_definition}
\begin{split}
    \text{RMSD}(\mathbf{x}_1,\mathbf{x}_2) ={}& 
    \sqrt{\frac{\sum_{i=1}^{N} \|\mathbf{x}_{1,i} - \mathbf{x}_{2,i} \|^2}{N}} \\
    ={}& \sqrt{\frac{\sum_{i=1}^{N} \sum_{j \in \{x,y,z\}} (x_{1,i,j} - x_{2,i,j})^2}{N}}
\end{split}
\end{equation}
with $N$ denoting the number of atoms and the index $i$ referring to the $i$-th atom, whose Cartesian coordinates (x,y,z) are included in the summation. Note that this definition differs from the one used in React-OT, where the RMSD is normalized by $3N$ instead. The mean reaction RMSD is calculated as the arithmetic mean of the RMSD values of the reactant, TS, and product structures relative to their reference counterparts. To capture more nuanced geometric errors, we introduce the bond distance mean absolute error (DMAE), defined as:
\begin{equation}\label{eq:dmae_definition}
    \text{DMAE}(\mathbf{x}_1,\mathbf{x}_2) = \frac{\sum_{i\not= j} |d_{1,ij} - d_{2,ij}|}{N(N-1)}
\end{equation}
with $d_{1,ij}=\|\mathbf{x}_{1,i} - \mathbf{x}_{1,j} \|^2$ as the interatomic distance between atom $i$ and $j$. To quantify deviations in reaction barriers, the electronic energies $V$ of all sampled TS structures is computed and the mean absolute error (MAE), defined as
\begin{equation}\label{eq:mae_definition}
\text{MAE} = \frac{1}{M} \sum_{i=1}^{M} \big| V(\mathbf{x}_i) - V(\hat{\mathbf{x}}_i) \big|,
\end{equation}
where $\hat{\mathbf{x}}_i$ denotes a predicted TS, $\mathbf{x}_i$ the corresponding reference TS from the dataset, and $M$ the number of samples. To assess how well the geometric statistics of the predictions match the reference data, the bond-length distributions are compared using the Wasserstein-1 distance. For two one-dimensional empirical distributions $p$ and $q$ with cumulative distribution functions $P$ and $Q$, respectively, this distance is given by
\begin{equation}
W_1(p,q) = \int_{-\infty}^{\infty} \lvert P(x) - Q(x) \rvert , \mathrm{d}x.
\end{equation}
The metric is evaluated independently for each bond type present in the dataset and then averaged over all types. Since the Wasserstein-1 distance measures the minimal cost required to morph one distribution into another, it provides a sensitive indicator of discrepancies in geometric features such as bond-length statistics.

\section*{Data and Software Availability}
The Transition1x dataset\cite{schreiner_transition1x_2022} used in this work is available via Figshare at \url{https://doi.org/10.6084/m9.figshare.19614657.v4}. The pretrained model checkpoints and the corresponding databases are available on Zenodo at \url{https://doi.org/10.5281/zenodo.18338077}. The codebase is publicly available as an open-source repository on GitHub to support continuous development at \url{https://github.com/samirdarouich/RobustTSGen}.

\section*{Acknowledgements}
This research was funded by the Ministry of Science, Research and the Arts Baden-Wuerttemberg in the Artificial Intelligence Software Academy (AISA). We thank the Deutsche Forschungsgemeinschaft (DFG, German Research Foundation) for supporting this work by funding - EXC2075 – 390740016 under Germany's Excellence Strategy. We acknowledge the support by the Stuttgart Center for Simulation Science (SimTech). H.J.K. was supported by a Simon Family Faculty Research Innovation Fund and an Alfred P. Sloan Fellowship in Chemistry. J.W.T. was partially supported by a Leslye Miller Fraser and Darryl M. Fraser Fellowship from the MIT School of Engineering. W.L. was partially supported by the MolSSI Software Fellowship. The authors acknowledge support by the state of Baden-Württemberg through bwHPC and the German Research Foundation (DFG) through grant no INST 40/575-1 FUGG (JUSTUS 2 cluster). We are grateful to Boshra Ariguib and Tanja Bien for her valuable input on improving the visual presentation of the manuscript. We thank Vinh Tong for helpful discussions during the development of the self-supervised pretraining strategy.


\begin{thebibliography}{10}
\expandafter\ifx\csname url\endcsname\relax
  \def\url#1{\texttt{#1}}\fi
\expandafter\ifx\csname urlprefix\endcsname\relax\def\urlprefix{URL }\fi
\providecommand{\bibinfo}[2]{#2}
\providecommand{\eprint}[2][]{\url{#2}}

\bibitem{banerjee_search_1985}
\bibinfo{author}{Banerjee, A.}, \bibinfo{author}{Adams, N.}, \bibinfo{author}{Simons, J.} \& \bibinfo{author}{Shepard, R.}
\newblock \bibinfo{title}{Search for stationary points on surfaces}.
\newblock \emph{\bibinfo{journal}{J. Phys. Chem.}} \textbf{\bibinfo{volume}{89}}, \bibinfo{pages}{52--57} (\bibinfo{year}{1985}).

\bibitem{baker_algorithm_1986}
\bibinfo{author}{Baker, J.}
\newblock \bibinfo{title}{An algorithm for the location of transition states}.
\newblock \emph{\bibinfo{journal}{J. Comput. Chem.}} \textbf{\bibinfo{volume}{7}}, \bibinfo{pages}{385--395} (\bibinfo{year}{1986}).

\bibitem{henkelman_dimer_1999}
\bibinfo{author}{Henkelman, G.} \& \bibinfo{author}{Jónsson, H.}
\newblock \bibinfo{title}{A dimer method for finding saddle points on high dimensional potential surfaces using only first derivatives}.
\newblock \emph{\bibinfo{journal}{J. Chem. Phys.}} \textbf{\bibinfo{volume}{111}}, \bibinfo{pages}{7010--7022} (\bibinfo{year}{1999}).

\bibitem{jonsson1998nudged}
\bibinfo{author}{J{\'o}nsson, H.}, \bibinfo{author}{Mills, G.} \& \bibinfo{author}{Jacobsen, K.~W.}
\newblock \bibinfo{title}{Nudged elastic band method for finding minimum energy paths of transitions}.
\newblock In \emph{\bibinfo{booktitle}{Classical and quantum dynamics in condensed phase simulations}}, \bibinfo{pages}{385--404} (\bibinfo{publisher}{World Scientific}, \bibinfo{year}{1998}).

\bibitem{henkelman_climbing_2000}
\bibinfo{author}{Henkelman, G.}, \bibinfo{author}{Uberuaga, B.~P.} \& \bibinfo{author}{Jónsson, H.}
\newblock \bibinfo{title}{A climbing image nudged elastic band method for finding saddle points and minimum energy paths}.
\newblock \emph{\bibinfo{journal}{J. Chem. Phys.}} \textbf{\bibinfo{volume}{113}}, \bibinfo{pages}{9901--9904} (\bibinfo{year}{2000}).

\bibitem{peters_growing_2004}
\bibinfo{author}{Peters, B.}, \bibinfo{author}{Heyden, A.}, \bibinfo{author}{Bell, A.~T.} \& \bibinfo{author}{Chakraborty, A.}
\newblock \bibinfo{title}{A growing string method for determining transition states: {Comparison} to the nudged elastic band and string methods}.
\newblock \emph{\bibinfo{journal}{J. Chem. Phys.}} \textbf{\bibinfo{volume}{120}}, \bibinfo{pages}{7877--7886} (\bibinfo{year}{2004}).

\bibitem{mardirossian2017thirty}
\bibinfo{author}{Mardirossian, N.} \& \bibinfo{author}{Head-Gordon, M.}
\newblock \bibinfo{title}{Thirty years of density functional theory in computational chemistry: an overview and extensive assessment of 200 density functionals}.
\newblock \emph{\bibinfo{journal}{Molecular physics}} \textbf{\bibinfo{volume}{115}}, \bibinfo{pages}{2315--2372} (\bibinfo{year}{2017}).

\bibitem{pozun_optimizing_2012}
\bibinfo{author}{Pozun, Z.~D.} \emph{et~al.}
\newblock \bibinfo{title}{Optimizing transition states via kernel-based machine learning}.
\newblock \emph{\bibinfo{journal}{J. Chem. Phys.}} \textbf{\bibinfo{volume}{136}}, \bibinfo{pages}{174101} (\bibinfo{year}{2012}).

\bibitem{koistinen_minimum_2016}
\bibinfo{author}{Koistinen, O.-P.}, \bibinfo{author}{Maras, E.}, \bibinfo{author}{Vehtari, A.} \& \bibinfo{author}{Jónsson, H.}
\newblock \bibinfo{title}{Minimum energy path calculations with gaussian process regression}.
\newblock \emph{\bibinfo{journal}{Nanosyst.:Phys., Chem., Math.}} \bibinfo{pages}{925–935} (\bibinfo{year}{2016}).

\bibitem{koistinen_nudged_2017}
\bibinfo{author}{Koistinen, O.-P.}, \bibinfo{author}{Dagbjartsdóttir, F.~B.}, \bibinfo{author}{Ásgeirsson, V.}, \bibinfo{author}{Vehtari, A.} \& \bibinfo{author}{Jónsson, H.}
\newblock \bibinfo{title}{Nudged elastic band calculations accelerated with {Gaussian} process regression}.
\newblock \emph{\bibinfo{journal}{J. Chem. Phys.}} \textbf{\bibinfo{volume}{147}}, \bibinfo{pages}{152720} (\bibinfo{year}{2017}).

\bibitem{denzel_gaussian_2018}
\bibinfo{author}{Denzel, A.} \& \bibinfo{author}{Kästner, J.}
\newblock \bibinfo{title}{Gaussian {Process} {Regression} for {Transition} {State} {Search}}.
\newblock \emph{\bibinfo{journal}{J. Chem. Theory Comput.}} \textbf{\bibinfo{volume}{14}}, \bibinfo{pages}{5777--5786} (\bibinfo{year}{2018}).

\bibitem{denzel_gaussian_2019}
\bibinfo{author}{Denzel, A.}, \bibinfo{author}{Haasdonk, B.} \& \bibinfo{author}{Kästner, J.}
\newblock \bibinfo{title}{Gaussian {Process} {Regression} for {Minimum} {Energy} {Path} {Optimization} and {Transition} {State} {Search}}.
\newblock \emph{\bibinfo{journal}{J. Phys. Chem. A}} \textbf{\bibinfo{volume}{123}}, \bibinfo{pages}{9600--9611} (\bibinfo{year}{2019}).

\bibitem{garrido_torres_low-scaling_2019}
\bibinfo{author}{Garrido~Torres, J.~A.}, \bibinfo{author}{Jennings, P.~C.}, \bibinfo{author}{Hansen, M.~H.}, \bibinfo{author}{Boes, J.~R.} \& \bibinfo{author}{Bligaard, T.}
\newblock \bibinfo{title}{Low-{Scaling} {Algorithm} for {Nudged} {Elastic} {Band} {Calculations} {Using} a {Surrogate} {Machine} {Learning} {Model}}.
\newblock \emph{\bibinfo{journal}{Phys. Rev. Lett.}} \textbf{\bibinfo{volume}{122}}, \bibinfo{pages}{156001} (\bibinfo{year}{2019}).

\bibitem{heinen_transition_2022}
\bibinfo{author}{Heinen, S.}, \bibinfo{author}{von Rudorff, G.~F.} \& \bibinfo{author}{von Lilienfeld, O.~A.}
\newblock \bibinfo{title}{Transition state search and geometry relaxation throughout chemical compound space with quantum machine learning}.
\newblock \emph{\bibinfo{journal}{J. Chem. Phys.}} \textbf{\bibinfo{volume}{157}}, \bibinfo{pages}{221102} (\bibinfo{year}{2022}).

\bibitem{peterson_acceleration_2016}
\bibinfo{author}{Peterson, A.~A.}
\newblock \bibinfo{title}{Acceleration of saddle-point searches with machine learning}.
\newblock \emph{\bibinfo{journal}{J. Chem. Phys.}} \textbf{\bibinfo{volume}{145}}, \bibinfo{pages}{074106} (\bibinfo{year}{2016}).

\bibitem{schreiner_neuralnebneural_2022}
\bibinfo{author}{Schreiner, M.}, \bibinfo{author}{Bhowmik, A.}, \bibinfo{author}{Vegge, T.}, \bibinfo{author}{Jørgensen, P.~B.} \& \bibinfo{author}{Winther, O.}
\newblock \bibinfo{title}{{NeuralNEB}—neural networks can find reaction paths fast}.
\newblock \emph{\bibinfo{journal}{Mach. Learn.: Sci. Technol}} \textbf{\bibinfo{volume}{3}}, \bibinfo{pages}{045022} (\bibinfo{year}{2022}).

\bibitem{zhang2024exploring}
\bibinfo{author}{Zhang, S.} \emph{et~al.}
\newblock \bibinfo{title}{Exploring the frontiers of condensed-phase chemistry with a general reactive machine learning potential}.
\newblock \emph{\bibinfo{journal}{Nat. Chem.}} \textbf{\bibinfo{volume}{16}}, \bibinfo{pages}{727--734} (\bibinfo{year}{2024}).

\bibitem{wander_cattsunami_2024}
\bibinfo{author}{Wander, B.}, \bibinfo{author}{Shuaibi, M.}, \bibinfo{author}{Kitchin, J.~R.}, \bibinfo{author}{Ulissi, Z.~W.} \& \bibinfo{author}{Zitnick, C.~L.}
\newblock \bibinfo{title}{Cattsunami: Accelerating transition state energy calculations with pretrained graph neural networks}.
\newblock \emph{\bibinfo{journal}{ACS Catal.}} \textbf{\bibinfo{volume}{15}}, \bibinfo{pages}{5283--5294} (\bibinfo{year}{2025}).

\bibitem{yuan_analytical_2024}
\bibinfo{author}{Yuan, E. C.-Y.} \emph{et~al.}
\newblock \bibinfo{title}{Analytical ab initio hessian from a deep learning potential for transition state optimization}.
\newblock \emph{\bibinfo{journal}{Nat. Commun.}} \textbf{\bibinfo{volume}{15}}, \bibinfo{pages}{8865} (\bibinfo{year}{2024}).

\bibitem{zhao_harnessing_2025}
\bibinfo{author}{Zhao, Q.} \emph{et~al.}
\newblock \bibinfo{title}{Harnessing machine learning to enhance transition state search with interatomic potentials and generative models}.
\newblock \emph{\bibinfo{journal}{Adv. Sci}} \textbf{\bibinfo{volume}{12}}, \bibinfo{pages}{e06240} (\bibinfo{year}{2025}).

\bibitem{yin2025cats}
\bibinfo{author}{Yin, J.} \emph{et~al.}
\newblock \bibinfo{title}{Cats: Toward scalable and efficient transition state screening for catalyst discovery}.
\newblock \emph{\bibinfo{journal}{ACS Catal.}} \textbf{\bibinfo{volume}{15}}, \bibinfo{pages}{15754--15764} (\bibinfo{year}{2025}).

\bibitem{tang2025accelerating}
\bibinfo{author}{Tang, K.} \emph{et~al.}
\newblock \bibinfo{title}{Accelerating transition state search and ligand screening for organometallic catalysis with reactive machine learning potential}.
\newblock \emph{\bibinfo{journal}{J. Chem. Theory Comput.}} \textbf{\bibinfo{volume}{21}}, \bibinfo{pages}{8933--8949} (\bibinfo{year}{2025}).

\bibitem{pattanaik_generating_2020}
\bibinfo{author}{Pattanaik, L.}, \bibinfo{author}{Ingraham, J.~B.}, \bibinfo{author}{Grambow, C.~A.} \& \bibinfo{author}{Green, W.~H.}
\newblock \bibinfo{title}{Generating transition states of isomerization reactions with deep learning}.
\newblock \emph{\bibinfo{journal}{Phys. Chem. Chem. Phys.}} \textbf{\bibinfo{volume}{22}}, \bibinfo{pages}{23618--23626} (\bibinfo{year}{2020}).

\bibitem{jackson_tsnet_2021}
\bibinfo{author}{Jackson, R.}, \bibinfo{author}{Zhang, W.} \& \bibinfo{author}{Pearson, J.}
\newblock \bibinfo{title}{{TSNet}: predicting transition state structures with tensor field networks and transfer learning}.
\newblock \emph{\bibinfo{journal}{Chem. Sci.}} \textbf{\bibinfo{volume}{12}}, \bibinfo{pages}{10022--10040} (\bibinfo{year}{2021}).

\bibitem{zhang_deep_2021}
\bibinfo{author}{Zhang, J.} \emph{et~al.}
\newblock \bibinfo{title}{Deep reinforcement learning of transition states}.
\newblock \emph{\bibinfo{journal}{Phys. Chem. Chem. Phys.}} \textbf{\bibinfo{volume}{23}}, \bibinfo{pages}{6888--6895} (\bibinfo{year}{2021}).

\bibitem{choi_prediction_2023}
\bibinfo{author}{Choi, S.}
\newblock \bibinfo{title}{Prediction of transition state structures of gas-phase chemical reactions via machine learning}.
\newblock \emph{\bibinfo{journal}{Nat. Commun.}} \textbf{\bibinfo{volume}{14}}, \bibinfo{pages}{1168} (\bibinfo{year}{2023}).

\bibitem{makos_generative_2021}
\bibinfo{author}{Makoś, M.~Z.}, \bibinfo{author}{Verma, N.}, \bibinfo{author}{Larson, E.~C.}, \bibinfo{author}{Freindorf, M.} \& \bibinfo{author}{Kraka, E.}
\newblock \bibinfo{title}{Generative adversarial networks for transition state geometry prediction}.
\newblock \emph{\bibinfo{journal}{J. Chem. Phys.}} \textbf{\bibinfo{volume}{155}}, \bibinfo{pages}{024116} (\bibinfo{year}{2021}).

\bibitem{duan_accurate_2023}
\bibinfo{author}{Duan, C.}, \bibinfo{author}{Du, Y.}, \bibinfo{author}{Jia, H.} \& \bibinfo{author}{Kulik, H.~J.}
\newblock \bibinfo{title}{Accurate transition state generation with an object-aware equivariant elementary reaction diffusion model}.
\newblock \emph{\bibinfo{journal}{Nat. Comput. Sci.}} \textbf{\bibinfo{volume}{3}}, \bibinfo{pages}{1045--1055} (\bibinfo{year}{2023}).

\bibitem{kim_diffusion-based_2024}
\bibinfo{author}{Kim, S.}, \bibinfo{author}{Woo, J.} \& \bibinfo{author}{Kim, W.~Y.}
\newblock \bibinfo{title}{Diffusion-based generative {AI} for exploring transition states from {2D} molecular graphs}.
\newblock \emph{\bibinfo{journal}{Nat. Commun.}} \textbf{\bibinfo{volume}{15}}, \bibinfo{pages}{341} (\bibinfo{year}{2024}).

\bibitem{galustian_goflow_2025}
\bibinfo{author}{Galustian, L.}, \bibinfo{author}{Mark, K.}, \bibinfo{author}{Karwounopoulos, J.}, \bibinfo{author}{Kovar, M. P.-P.} \& \bibinfo{author}{Heid, E.}
\newblock \bibinfo{title}{Goflow: Efficient transition state geometry prediction with flow matching and e (3)-equivariant neural networks}.
\newblock \emph{\bibinfo{journal}{chemrxiv preprint chemrxiv-2025-bk2rh}}  (\bibinfo{year}{2025}).

\bibitem{duan_optimal_2025}
\bibinfo{author}{Duan, C.} \emph{et~al.}
\newblock \bibinfo{title}{Optimal transport for generating transition states in chemical reactions}.
\newblock \emph{\bibinfo{journal}{Nat. Mach. Intell.}} \textbf{\bibinfo{volume}{7}}, \bibinfo{pages}{615--626} (\bibinfo{year}{2025}).

\bibitem{hayashi_generative_2025}
\bibinfo{author}{Hayashi, A.}, \bibinfo{author}{Takamoto, S.}, \bibinfo{author}{Li, J.}, \bibinfo{author}{Tsuboi, Y.} \& \bibinfo{author}{Okanohara, D.}
\newblock \bibinfo{title}{Generative {Model} for {Constructing} {Reaction} {Path} from {Initial} to {Final} {States}}.
\newblock \emph{\bibinfo{journal}{J. Chem. Theory Comput.}} \textbf{\bibinfo{volume}{21}}, \bibinfo{pages}{1292--1305} (\bibinfo{year}{2025}).

\bibitem{darouich2025adaptive}
\bibinfo{author}{Darouich, S.}, \bibinfo{author}{Tong, V.}, \bibinfo{author}{Bien, T.}, \bibinfo{author}{K{\"a}stner, J.} \& \bibinfo{author}{Niepert, M.}
\newblock \bibinfo{title}{Adaptive transition state refinement with learned equilibrium flows}.
\newblock \emph{\bibinfo{journal}{arXiv preprint arXiv:2507.16521}}  (\bibinfo{year}{2025}).

\bibitem{ho2020denoising}
\bibinfo{author}{Ho, J.}, \bibinfo{author}{Jain, A.} \& \bibinfo{author}{Abbeel, P.}
\newblock \bibinfo{title}{Denoising diffusion probabilistic models}.
\newblock \emph{\bibinfo{journal}{Advances in neural information processing systems}} \textbf{\bibinfo{volume}{33}}, \bibinfo{pages}{6840--6851} (\bibinfo{year}{2020}).

\bibitem{lipman_flow_2022}
\bibinfo{author}{Lipman, Y.}, \bibinfo{author}{Chen, R.~T.}, \bibinfo{author}{Ben-Hamu, H.}, \bibinfo{author}{Nickel, M.} \& \bibinfo{author}{Le, M.}
\newblock \bibinfo{title}{Flow matching for generative modeling}.
\newblock \emph{\bibinfo{journal}{arXiv preprint arXiv:2210.02747}}  (\bibinfo{year}{2022}).

\bibitem{liu_flow_2022}
\bibinfo{author}{Liu, X.}, \bibinfo{author}{Gong, C.} \& \bibinfo{author}{Liu, Q.}
\newblock \bibinfo{title}{Flow straight and fast: Learning to generate and transfer data with rectified flow}.
\newblock \emph{\bibinfo{journal}{arXiv preprint arXiv:2209.03003}}  (\bibinfo{year}{2022}).

\bibitem{mark2025feynman}
\bibinfo{author}{Mark, K.}, \bibinfo{author}{Galustian, L.}, \bibinfo{author}{Kovar, M. P.-P.} \& \bibinfo{author}{Heid, E.}
\newblock \bibinfo{title}{Feynman-kac-flow: Inference steering of conditional flow matching to an energy-tilted posterior}.
\newblock \emph{\bibinfo{journal}{arXiv preprint arXiv:2509.01543}}  (\bibinfo{year}{2025}).

\bibitem{si2025transition}
\bibinfo{author}{Si, Y.} \emph{et~al.}
\newblock \bibinfo{title}{Transition state structure detection with machine learning{\'s}}.
\newblock \emph{\bibinfo{journal}{npj Comput. Mater.}} \textbf{\bibinfo{volume}{11}}, \bibinfo{pages}{199} (\bibinfo{year}{2025}).

\bibitem{schreiner_transition1x_2022}
\bibinfo{author}{Schreiner, M.}, \bibinfo{author}{Bhowmik, A.}, \bibinfo{author}{Vegge, T.}, \bibinfo{author}{Busk, J.} \& \bibinfo{author}{Winther, O.}
\newblock \bibinfo{title}{Transition1x - a dataset for building generalizable reactive machine learning potentials}.
\newblock \emph{\bibinfo{journal}{Sci. Data}} \textbf{\bibinfo{volume}{9}}, \bibinfo{pages}{779} (\bibinfo{year}{2022}).

\bibitem{grambow2020reactants}
\bibinfo{author}{Grambow, C.~A.}, \bibinfo{author}{Pattanaik, L.} \& \bibinfo{author}{Green, W.~H.}
\newblock \bibinfo{title}{Reactants, products, and transition states of elementary chemical reactions based on quantum chemistry}.
\newblock \emph{\bibinfo{journal}{Sci. Data}} \textbf{\bibinfo{volume}{7}}, \bibinfo{pages}{137} (\bibinfo{year}{2020}).

\bibitem{nandy_computational_2021}
\bibinfo{author}{Nandy, A.} \emph{et~al.}
\newblock \bibinfo{title}{Computational {Discovery} of {Transition}-metal {Complexes}: {From} {High}-throughput {Screening} to {Machine} {Learning}}.
\newblock \emph{\bibinfo{journal}{Chem. Rev.}} \textbf{\bibinfo{volume}{121}}, \bibinfo{pages}{9927--10000} (\bibinfo{year}{2021}).

\bibitem{chen2022automated}
\bibinfo{author}{Chen, S.} \emph{et~al.}
\newblock \bibinfo{title}{Automated construction and optimization combined with machine learning to generate pt (ii) methane c--h activation transition states}.
\newblock \emph{\bibinfo{journal}{Topics in Catalysis}} \textbf{\bibinfo{volume}{65}}, \bibinfo{pages}{312--324} (\bibinfo{year}{2022}).

\bibitem{toney2025exploring}
\bibinfo{author}{Toney, J.~W.} \emph{et~al.}
\newblock \bibinfo{title}{Exploring beyond experiment: generating high-quality datasets of transition metal complexes with quantum chemistry and machine learning}.
\newblock \emph{\bibinfo{journal}{Curr. Opin. Chem. Eng.}} \textbf{\bibinfo{volume}{50}}, \bibinfo{pages}{101189} (\bibinfo{year}{2025}).

\bibitem{li2021computational}
\bibinfo{author}{Li, W.}, \bibinfo{author}{Ma, H.}, \bibinfo{author}{Li, S.} \& \bibinfo{author}{Ma, J.}
\newblock \bibinfo{title}{Computational and data driven molecular material design assisted by low scaling quantum mechanics calculations and machine learning}.
\newblock \emph{\bibinfo{journal}{Chem. Sci.}} \textbf{\bibinfo{volume}{12}}, \bibinfo{pages}{14987--15006} (\bibinfo{year}{2021}).

\bibitem{Bannwarth2021}
\bibinfo{author}{Bannwarth, C.}, \bibinfo{author}{Ehlert, S.} \& \bibinfo{author}{Grimme, S.}
\newblock \bibinfo{title}{Gfn2-xtb—an accurate and broadly parametrized self-consistent tight-binding quantum chemical method with multipole electrostatics and density-dependent dispersion contributions}.
\newblock \emph{\bibinfo{journal}{J. Chem. Theory Comput.}} \textbf{\bibinfo{volume}{15}}, \bibinfo{pages}{1652--1671} (\bibinfo{year}{2019}).

\bibitem{peng_moldiff_2023}
\bibinfo{author}{Peng, X.}, \bibinfo{author}{Guan, J.}, \bibinfo{author}{Liu, Q.} \& \bibinfo{author}{Ma, J.}
\newblock \bibinfo{title}{Moldiff: Addressing the atom-bond inconsistency problem in 3d molecule diffusion generation}.
\newblock \emph{\bibinfo{journal}{arXiv preprint arXiv:2305.07508}}  (\bibinfo{year}{2023}).

\bibitem{williams_physics-informed_2024}
\bibinfo{author}{Williams, D.~C.} \& \bibinfo{author}{Inala, N.}
\newblock \bibinfo{title}{Physics-{Informed} {Generative} {Model} for {Drug}-like {Molecule} {Conformers}}.
\newblock \emph{\bibinfo{journal}{J. Chem. Inf. Model.}} \textbf{\bibinfo{volume}{64}}, \bibinfo{pages}{2988--3007} (\bibinfo{year}{2024}).

\bibitem{vost_improving_2025}
\bibinfo{author}{Vost, L.}, \bibinfo{author}{Chenthamarakshan, V.}, \bibinfo{author}{Das, P.} \& \bibinfo{author}{Deane, C.~M.}
\newblock \bibinfo{title}{Improving structural plausibility in diffusion-based {3D} molecule generation via property-conditioned training with distorted molecules}.
\newblock \emph{\bibinfo{journal}{Digital Discovery}} \textbf{\bibinfo{volume}{4}}, \bibinfo{pages}{1092--1099} (\bibinfo{year}{2025}).

\bibitem{wohlwend_boltz-1_2025}
\bibinfo{author}{Wohlwend, J.} \emph{et~al.}
\newblock \bibinfo{title}{Boltz-1 democratizing biomolecular interaction modeling}.
\newblock \emph{\bibinfo{journal}{BioRxiv}} \bibinfo{pages}{2024--11} (\bibinfo{year}{2025}).

\bibitem{fukui1981path}
\bibinfo{author}{Fukui, K.}
\newblock \bibinfo{title}{The path of chemical reactions-the irc approach}.
\newblock \emph{\bibinfo{journal}{Acc. Chem. Res.}} \textbf{\bibinfo{volume}{14}}, \bibinfo{pages}{363--368} (\bibinfo{year}{1981}).

\bibitem{groom2016CSD}
\bibinfo{author}{Groom, C.~R.}, \bibinfo{author}{Bruno, I.~J.}, \bibinfo{author}{Lightfoot, M. P.~L.} \& \bibinfo{author}{Ward, S.~C.}
\newblock \bibinfo{title}{The cambridge structural database}.
\newblock \emph{\bibinfo{journal}{Acta Crystallogr. B, Struct. Sci. Cryst. Eng. Mater.}} \textbf{\bibinfo{volume}{72}}, \bibinfo{pages}{171--179} (\bibinfo{year}{2016}).

\bibitem{weininger1988SMILES}
\bibinfo{author}{Weininger, D.}
\newblock \bibinfo{title}{Smiles, a chemical language and information system}.
\newblock \emph{\bibinfo{journal}{J. Chem. Inf. Comput.}} \textbf{\bibinfo{volume}{28}}, \bibinfo{pages}{31--36} (\bibinfo{year}{1988}).

\bibitem{toney2025pydentate1}
\bibinfo{author}{Toney, J.~W.}, \bibinfo{author}{St.~Michel, R.~G.}, \bibinfo{author}{Garrison, A.~G.}, \bibinfo{author}{Kevlishvili, I.} \& \bibinfo{author}{Kulik, H.~J.}
\newblock \bibinfo{title}{Graph neural networks for predicting metal–ligand coordination of transition metal complexes}.
\newblock \emph{\bibinfo{journal}{Proc. Natl. Acad. Sci.}} \textbf{\bibinfo{volume}{122}}, \bibinfo{pages}{e2415658122} (\bibinfo{year}{2025}).

\bibitem{toney2025pydentate2}
\bibinfo{author}{Toney, J.~W.}, \bibinfo{author}{St.~Michel, R.~G.}, \bibinfo{author}{Garrison, A.~G.}, \bibinfo{author}{Kevlishvili, I.} \& \bibinfo{author}{Kulik, H.~J.}
\newblock \bibinfo{title}{Identifying dynamic metal–ligand coordination modes with ensemble learning}.
\newblock \emph{\bibinfo{journal}{J. Am. Chem. Soc.}}  (\bibinfo{year}{2025}).

\bibitem{molSimplify}
\bibinfo{author}{Ioannidis, E.~I.}, \bibinfo{author}{Gani, T. Z.~H.} \& \bibinfo{author}{Kulik, H.~J.}
\newblock \bibinfo{title}{molsimplify: A toolkit for automating discovery in inorganic chemistry}.
\newblock \emph{\bibinfo{journal}{J. Comput. Chem.}} \textbf{\bibinfo{volume}{37}}, \bibinfo{pages}{2106--2117} (\bibinfo{year}{2016}).

\bibitem{terrones2025molsimplify}
\bibinfo{author}{Terrones, G.} \emph{et~al.}
\newblock \bibinfo{title}{molsimplify 2.0: Improved structure generation for automating discovery in inorganic molecular and reticular chemistry}.
\newblock \emph{\bibinfo{journal}{ChemRxiv}}  (\bibinfo{year}{2025}).

\bibitem{ase-paper}
\bibinfo{author}{Larsen, A.~H.} \emph{et~al.}
\newblock \bibinfo{title}{The atomic simulation environment—a python library for working with atoms}.
\newblock \emph{\bibinfo{journal}{J. Phys.:Condens. Matter}} \textbf{\bibinfo{volume}{29}}, \bibinfo{pages}{273002} (\bibinfo{year}{2017}).

\bibitem{hermes2022}
\bibinfo{author}{Hermes, E.~D.}, \bibinfo{author}{Sargsyan, K.}, \bibinfo{author}{Najm, H.~N.} \& \bibinfo{author}{Z{\'a}dor, J.}
\newblock \bibinfo{title}{Sella, an open-source automation-friendly molecular saddle point optimizer}.
\newblock \emph{\bibinfo{journal}{J. Chem. Theory Comput.}} \textbf{\bibinfo{volume}{18}}, \bibinfo{pages}{6974--6988} (\bibinfo{year}{2022}).

\bibitem{ORCA}
\bibinfo{author}{Neese, F.}
\newblock \bibinfo{title}{The orca program system}.
\newblock \emph{\bibinfo{journal}{WIRES Comput. Molec. Sci.}} \textbf{\bibinfo{volume}{2}}, \bibinfo{pages}{73--78} (\bibinfo{year}{2012}).

\bibitem{chai2008systematic}
\bibinfo{author}{Chai, J.-D.} \& \bibinfo{author}{Head-Gordon, M.}
\newblock \bibinfo{title}{Systematic optimization of long-range corrected hybrid density functionals}.
\newblock \emph{\bibinfo{journal}{J. Chem. Phys.}} \textbf{\bibinfo{volume}{128}} (\bibinfo{year}{2008}).

\bibitem{weigend2005balanced}
\bibinfo{author}{Weigend, F.} \& \bibinfo{author}{Ahlrichs, R.}
\newblock \bibinfo{title}{Balanced basis sets of split valence, triple zeta valence and quadruple zeta valence quality for h to rn: Design and assessment of accuracy}.
\newblock \emph{\bibinfo{journal}{Phys. Chem. Chem. Phys.}} \textbf{\bibinfo{volume}{7}}, \bibinfo{pages}{3297--3305} (\bibinfo{year}{2005}).

\bibitem{grimme2010consistent}
\bibinfo{author}{Grimme, S.}, \bibinfo{author}{Antony, J.}, \bibinfo{author}{Ehrlich, S.} \& \bibinfo{author}{Krieg, H.}
\newblock \bibinfo{title}{A consistent and accurate ab initio parametrization of density functional dispersion correction (dft-d) for the 94 elements h-pu}.
\newblock \emph{\bibinfo{journal}{J. Chem. Phys.}} \textbf{\bibinfo{volume}{132}} (\bibinfo{year}{2010}).

\bibitem{adamo1999toward}
\bibinfo{author}{Adamo, C.} \& \bibinfo{author}{Barone, V.}
\newblock \bibinfo{title}{Toward reliable density functional methods without adjustable parameters: The pbe0 model}.
\newblock \emph{\bibinfo{journal}{J. Chem. Phys.}} \textbf{\bibinfo{volume}{110}}, \bibinfo{pages}{6158--6170} (\bibinfo{year}{1999}).

\bibitem{becke2005density}
\bibinfo{author}{Becke, A.~D.} \& \bibinfo{author}{Johnson, E.~R.}
\newblock \bibinfo{title}{A density-functional model of the dispersion interaction}.
\newblock \emph{\bibinfo{journal}{J. Chem. Phys.}} \textbf{\bibinfo{volume}{123}} (\bibinfo{year}{2005}).

\bibitem{pracht2020automated}
\bibinfo{author}{Pracht, P.}, \bibinfo{author}{Bohle, F.} \& \bibinfo{author}{Grimme, S.}
\newblock \bibinfo{title}{Automated exploration of the low-energy chemical space with fast quantum chemical methods}.
\newblock \emph{\bibinfo{journal}{Phys. Chem. Chem. Phys.}} \textbf{\bibinfo{volume}{22}}, \bibinfo{pages}{7169--7192} (\bibinfo{year}{2020}).

\bibitem{pracht2024crest}
\bibinfo{author}{Pracht, P.} \emph{et~al.}
\newblock \bibinfo{title}{Crest—a program for the exploration of low-energy molecular chemical space}.
\newblock \emph{\bibinfo{journal}{J. Chem. Phys.}} \textbf{\bibinfo{volume}{160}} (\bibinfo{year}{2024}).

\bibitem{spicher2020robust}
\bibinfo{author}{Spicher, S.} \& \bibinfo{author}{Grimme, S.}
\newblock \bibinfo{title}{Robust atomistic modeling of materials, organometallic, and biochemical systems}.
\newblock \emph{\bibinfo{journal}{Angew. Chem., Int. Ed.}} \textbf{\bibinfo{volume}{59}}, \bibinfo{pages}{15665--15673} (\bibinfo{year}{2020}).

\end{thebibliography}
\end{document}

\maketitle

\tableofcontents

\section{Model}

\subsection{Architecture}

The LEFTNet\cite{du2023new} architecture utilized in the original React-OT\cite{duan_optimal_2025} paper uses a one-hot encondig for each H, C, N, O atom type and additionally passes the atomic number as feature. These are encoded through a multi-layer perceptron (MLP) and then passed to the backbone LEFTNet architecture. In order to test React-OT on a diverser set of elements, a lookup embedding for each atomic number as well as its electronic properties is instead used in the combination with a MLP projection. To train these new atomic embeddings, the OAReactDiff\cite{duan_accurate_2023} backbone from the published checkpoint is frozen, and the new embeddings are trained for 200 epochs. This new checkpoint is then later on used for any downstream task. In order to verify the same expressiveness, a vanilla React-OT model on the original Transition1x\cite{schreiner_transition1x_2022} dataset is finetuned with the new embedding, leading to a median structural RMSD of 0.096~\AA\ and a median energetic difference of 1.123~kcal~mol$^{-1}$. For comparison, the original React-OT model achieves a median structural RMSD of 0.092~\AA\ and a median energetic difference of 1.092~kcal~mol$^{-1}$

During the testing of the vanilla LEFTNet architecture on the transition metal complex (TMC) and catalytic systems, it came to numerical overflows for bigger molecules, resulting in unstable predictions. The default LEFTNet architecture used in React-OT uses a sum aggregation along with a fully connected graph between reactant, TS, and product, leading to 22350 edges for a 50 atom reaction. In order to avoid this behavior, the newly trained embedding is frozen, and the aggregation type in the LEFTNet architecture is changed to mean instead of sum. The OAReactDiff model is further finetuned for 200 epochs. Again, the performance of the model was verified on the original Transition1x and lead to a median structural RMSD of 0.099~\AA\ and a median energetic difference of 1.232~kcal~mol$^{-1}$. This checkpoint is used for all systems including TMC's, while the original sum aggregation is used for the \swap benchmark.

For Adaptive Equilibrium Flow Matching (\aefm) \cite{darouich2025adaptive}, we employ the Equiformer-V2 architecture as used in the original work. Since \aefm is applied as a standalone refinement model, the bond loss weight is reduced to 0.5 to improve robustness for TS guesses that deviate more strongly from the target structure. The initialization of \aefm is identical to that used for React-OT, namely a linear interpolation between reactant and product coordinates. In contrast to React-OT, \aefm is not conditioned on reactants and products, which reduces computational overhead.

\subsection{Training pipeline}

For all experiments including React-OT the newly adapted OAReactDiff checkpoints are fine-tuned using the React-OT flow matching (FM) loss for 200 epochs using a learning rate of 2.5e-4. The vanilla React-OT model, is fine-tuned on the newly created \xtb dataset, in order to have a benchmark on HCNO chemistry. The CREST pretraining is done by taking the newly adapted OAReactDiff checkpoint and further training it for 200 epochs and a learning rate of 2.5e-4 with the pseudo-reactions. This checkpoint is then subsequently refined with the desired target data and the React-OT FM loss.

For \aefm, a similar training pipeline is employed. An initial model is trained on Transition1x and subsequently fine-tuned on \xtb to obtain the vanilla baseline. For the \swap- and \tmc-fine-tuned variants, the Transition1x-trained \aefm model serves as the initialization. All \aefm training runs are terminated using early stopping, with an additional 10\% of the training data held out for validation.

\section{Datasets}

The \swap dataset comprises not only reactions generated through atomic substitutions, but also the corresponding source reactions from Transition1x optimized at the GFN2-xTB level. Including the original HCNO reactions in the training set improves the performance of the fine-tuned React-OT model, yielding a median RMSD of 0.14~\AA\ and a median energetic deviation of 2.97~kcal~mol$^{-1}$, compared to a model fine-tuned exclusively on swap type 1 and 2 reactions, which attains a median RMSD of 0.17~\AA\ and a median energetic deviation of 4.75~kcal~mol$^{-1}$. This improvement can be attributed to the increased training set size, which comprises 18594 reactions when including both \xtb and swapped reactions, compared to 12396 reactions when only swapped reactions are used. This effect is not observed for the \tmc dataset, which already contains more than 36000 reactions. A React-OT model trained on 32633 \tmc reactions achieves a median RMSD of 0.24~\AA\ and a median energetic deviation of 6.68~kcal~mol$^{-1}$. Adding 6198 \xtb reactions to the training set does not alter performance, as the resulting model attains identical median RMSD and energetic deviation values. This indicates that, in this regime, additional organic reactions do not further influence model behavior. Based on these observations, all \swap-trained models include both the swapped and organic reactions, whereas \tmc-trained models are trained exclusively on TMC-substituted reactions.

\begin{figure}[ht]
\centering
\includegraphics[width=0.9\textwidth]{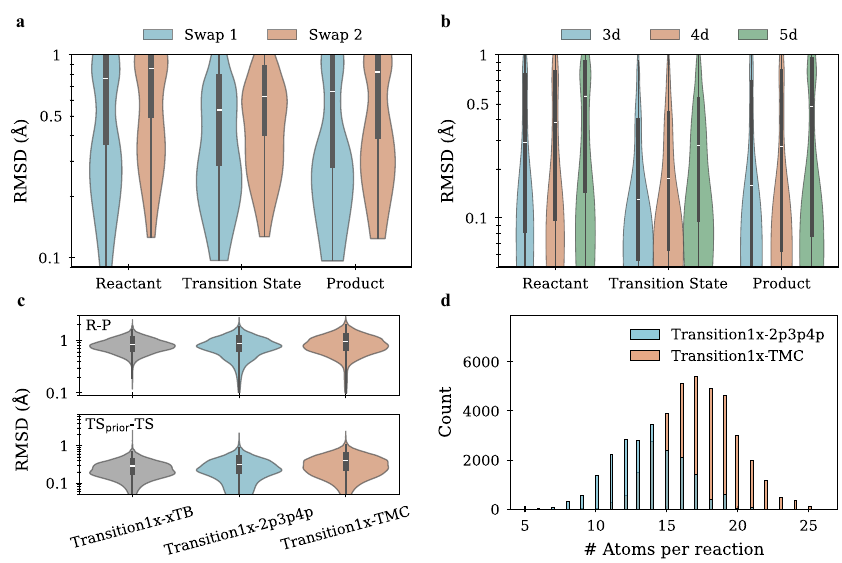}
\caption{
\textbf{Statistics for \swap and \tmc.}
\textbf{a} and \textbf{b} Shown is the RMSD of the reactant, TS, and product structures of the \swap and \tmc datasets with respect to their corresponding source structures from \xtb.
\textbf{c} RMSD distributions for reactant–to–product structures, as well as the prior TS guess used in React-OT to the true TS structures, for \xtb, \swap, and \tmc.
\textbf{d} Distribution of the number of atoms participating in the reactions for \swap and \tmc.
}\label{fig:sup_dataset_stats}
\end{figure}

As both \swap and \tmc are derived from Transition1x, it is instructive to quantify how their reaction trajectories deviate from the original \xtb references (Supplementary Figure~\ref{fig:sup_dataset_stats}a). For the computation of RMSDs between \xtb and \tmc structures, the added TMC substituent is removed prior to alignment so that only the common organic part is compared. A pronounced deviation is observed for the reactant and product structures in \swap. To elucidate this effect, we further analyze the number of formed and broken bonds in \swap relative to \xtb (Supplementary Table~\ref{tab:sup_reactive_bonds}). Approximately 40\% of reactions involve the swapped atom directly, and both \swap subsets exhibit a globally reduced reactivity, consistent with the stabilizing influence of heavier substituents.

For \tmc, deviations from the \xtb reference increase systematically with the period of the substituted TMC atom (Supplementary Figure~\ref{fig:sup_dataset_stats}b). As an additional structural analysis, Supplementary Figure~\ref{fig:sup_dataset_stats}c compares (i) the RMSD between reactant and product, and (ii) the RMSD between the linear prior and the target TS structure across the \xtb, \swap, and \tmc datasets. The reactant-product RMSD distributions for \swap and \tmc both exhibit extended low-RMSD tails, indicating a larger proportion of reactions involving relatively minor structural rearrangements. In contrast, the RMSD between the linear prior and the TS structure reflects the intrinsic difficulty of the dataset. Here, \tmc yields the largest deviations, confirming that it represents the most challenging setting, with the linear prior lying farthest from the true TS and thereby posing a more demanding learning problem. Finally, Supplementary Figure~\ref{fig:sup_dataset_stats}d shows the distribution of the number of atoms participating in each reaction, highlighting that \tmc consists of larger molecular systems with correspondingly more degrees of freedom.

\begin{table*}[ht]
\centering
\caption{Reactivity metrics for \xtb and \swap.}
\vspace{0.5em}
\label{tab:sup_reactive_bonds}
\begin{tabular}{@{\extracolsep\fill}lllll}
\toprule
Type & \# Reactive atoms & \# Formed bonds & \# Broken bonds & Swapped atom in \\
& & & &  reactive center (\%) \\
\midrule
\xtb & 3.52 & 1.13 & 1.54 & - \\
Swap 1 & 2.23 & 0.72 & 0.97 & 40.6 \\
Swap 2 & 2.19 & 0.77 & 0.92 & 42.2 \\
\bottomrule
\end{tabular}
\end{table*}

GFN2-xTB provides a practical balance between computational efficiency and accuracy. Supplementary Table~\ref{tab:sup_timing} reports the average wall time required to process one reaction, including both the P-RFO optimization and the subsequent IRC calculations. A \tmc reaction is found to require nearly three times the computational time of a \swap reaction, which can be attributed to the larger molecular sizes involved (\swap: 13.5 atoms on average; \tmc: 17.3 atoms on average). For reference, Supplementary Table~\ref{tab:sup_timing} also includes DFT timings for a set of 100 \tmc reactions, illustrating that GFN2-xTB is more than 700-fold faster than the corresponding DFT calculations.

\begin{table*}[ht]
\centering
\caption{Timing for dataset creation for \swap and \tmc using P-RFO and IRC.}
\vspace{0.5em}
\label{tab:sup_timing}
\begin{tabular}{@{\extracolsep\fill}lll}
\toprule
Type & Dataset & Time per reaction (s)\\
\midrule
GFN2-xTB & \swap & 18.3 \\
GFN2-xTB & \tmc & 46.5 \\
DFT & \tmc & 33174.7 \\
\bottomrule
\end{tabular}
\end{table*}

\section{Additional analysis}

\subsection{Similar reactions}

To assess the transferability of React-OT to previously unseen elements while preserving comparable reaction mechanisms, we identify structurally similar reactions between the altered datasets, \swap or \tmc, and their corresponding \xtb source reactions. A reaction is considered structurally similar if, for each structure along the reaction path, reactant, TS, and product, the RMSD to the original \xtb structure is below 0.2~\AA. Under this criterion, 40 reactions from \swap qualify as similar, comprising 36 swap type 1 reactions and 4 swap type 2 reactions. For \tmc, 197 distinct \xtb reactions have at least one structurally similar \tmc counterpart. However, not every TM leads to the same set of similar reactions, these 197 source reactions give rise to a total of 928 \tmc reactions, with the number of similar reactions varying across metals. Supplementary Table~\ref{tab:sup_similar_reactions} summarizes the performance of the vanilla React-OT model on the structurally similar \swap and \tmc reactions, alongside its performance on the corresponding \xtb source reactions. The pseudo-reaction pretrained React-OT models are also evaluated, demonstrating improved accuracy in both \swap and \tmc settings.

\begin{table*}[ht]
\centering
\caption{Statistic on similar reactions.}
\vspace{0.5em}
\label{tab:sup_similar_reactions}
\begin{adjustbox}{max width=\linewidth}
\begin{tabular}{@{\extracolsep\fill}llllllllll}
\toprule
Approach & Dataset & \multicolumn{2}{c}{RMSD (\AA)} & \multicolumn{2}{c}{DMAE (\AA)} & \multicolumn{2}{c}{$|\Delta E_{\text{TS}}|$ (kcal mol$^{-1}$)} & \# Reactions \\
\cmidrule(lr){3-4} \cmidrule(lr){5-6} \cmidrule(lr){7-8}
& & Mean & Median & Mean & Median & Mean & Median & \\
\midrule
Vanilla React-OT & \xtb & 0.06 & 0.04 & 0.03 & 0.02 & 1.19 & 0.54 & 40 \\
Vanilla React-OT & \swap & 0.21 & 0.18 & 0.11 & 0.09 & 177.51 & 130.66 & 40 \\
Vanilla React-OT\textsuperscript{\emph{a}} & \swap & 0.11 & 0.10 & 0.06 & 0.06 & 21.43 & 18.21 & 40 \\
Vanilla React-OT & \xtb & 0.08 & 0.05 & 0.12 & 0.10 & 1.74 & 0.65 & 197 \\
Vanilla React-OT & \tmc & 0.42 & 0.39 & 0.27 & 0.26 & 2396.83 & 1648.49 & 928 \\
Vanilla React-OT\textsuperscript{\emph{a}} & \tmc & 0.26 & 0.19 & 0.12 & 0.10 & 18.96 & 10.43 & 928 \\
\bottomrule
\end{tabular}
\end{adjustbox}
\footnotesize
\justify
\textsuperscript{\emph{a}} Pretraining with 2500 pseudo-reactions of the corresponding dataset. \\
\end{table*}

\subsection{Behavior on novel elements}

To obtain a more detailed geometric assessment, we analyze the Wasserstein-1 distance between the empirical bonded-distance distributions and those generated by React-OT. While React-OT predicts bond distances between familiar organic elements with good accuracy, bonds involving previously unseen elements are often assigned lengths similar to their organic analogues, resulting in broad and frequently unphysical distributions with a tendency toward unrealistically short values. Supplementary Table~\ref{tab:sup_wasserstein1d} summarizes these distances for the vanilla, fine-tuned, and self-supervised pretrained React-OT models. A clear separation is observed between HCNO-only bonds, which yield consistently low Wasserstein-1 distances, and bonds involving novel elements. This behavior is further illustrated in Supplementary Figure~\ref{fig:sup_bond_distributions}. Panel (a) compares the empirical bonded-distance distributions with those predicted by the vanilla React-OT model for substituted carbon and nitrogen derivatives in the \swap dataset, while panels (b) and (c) show the corresponding distributions for bonds involving carbon and nitrogen in the \tmc dataset. Across all cases, React-OT accurately reproduces the bonded-distance distributions for bond types present in the \xtb reference data. In contrast, for bond types absent from the training set, the model exhibits pronounced uncertainty, generating broad and frequently unphysical distance distributions that extend toward unrealistically short bond lengths. 

A post hoc relaxation of React-OT–predicted structures, in which atomic positions are adjusted to match the mean bond lengths of the corresponding bond types observed in the dataset, reinforces this conclusion by showing that most extreme errors originate from these unphysical distances. For the \swap benchmark, the median Wasserstein-1 distance decreases from 0.29~\AA\ to 0.07~\AA. A similar reduction is seen for the \tmc benchmark, where the median Wasserstein-1 distance drops from 0.30~\AA\ to 0.07~\AA\ (Supplementary Table~\ref{tab:sup_wasserstein1d}). Although adjusting bond lengths to match the dataset mean significantly improves local bond statistics, it only marginally affects overall structural accuracy and does not resolve the fundamental limitations of the predictions.

\begin{table*}[ht]
\centering
\caption{Wasserstein-1 distance between the empirical and the model-generated bond distributions.}
\vspace{0.5em}
\label{tab:sup_wasserstein1d}
\begin{tabular}{@{\extracolsep\fill}llllll}
\toprule
Approach & Dataset & \multicolumn{2}{c}{Wasserstein-1 distance} & \multicolumn{2}{c}{Wasserstein-1 distance} \\
& & \multicolumn{2}{c}{all elements (\AA)} & \multicolumn{2}{c}{HCNO elements (\AA)} \\
\cmidrule(lr){3-4} \cmidrule(lr){5-6}
& & Mean & Median & Mean & Median \\
\midrule
Vanilla React-OT & \swap & 0.32 & 0.29 & 0.04 & 0.01 \\
Vanilla React-OT\textsuperscript{\emph{a}} & \swap & 0.22 & 0.16 & 0.03 & 0.01 \\
Vanilla React-OT\textsuperscript{\emph{b}} & \swap & 0.07 & 0.07 & 0.06 & 0.06 \\
Vanilla React-OT & \tmc & 0.33 & 0.30 & 0.03 & 0.02 \\
Vanilla React-OT\textsuperscript{\emph{a}} & \tmc & 0.14 & 0.10 & 0.02 & 0.01 \\
Vanilla React-OT\textsuperscript{\emph{b}} & \tmc & 0.07 & 0.07 & 0.07 & 0.07 \\
Fine-tuned React-OT & \swap & 0.12 & 0.02 & 0.03 & 0.01 \\
Fine-tuned React-OT\textsuperscript{\emph{a}} & \swap & 0.12 & 0.02 & 0.02 & 0.01 \\
Fine-tuned React-OT & \tmc & 0.08 & 0.04 & 0.02 & 0.01\\
Fine-tuned React-OT\textsuperscript{\emph{a}} & \tmc & 0.07 & 0.03 & 0.02 & 0.01\\
\bottomrule
\end{tabular}
\footnotesize
\justify
\textsuperscript{\emph{a}} Pretraining with 2500 pseudo-reactions of the corresponding dataset. \\
\textsuperscript{\emph{b}} Post hoc relaxation to empirical mean bond length. \\
\end{table*}

\begin{figure}[ht]
\centering
\includegraphics[width=1.0\textwidth]{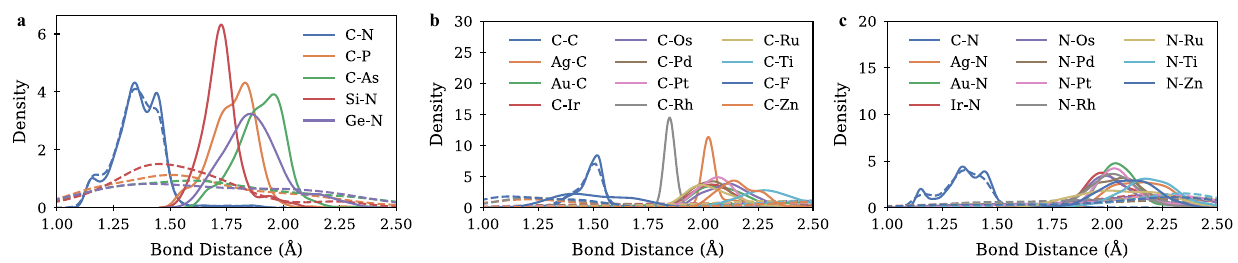}
\caption{
\textbf{Comparison of empirical (solid) and generated (dashed) bond distributions for vanilla React-OT model.} 
\textbf{a} Bond distributions for swapped derivatives of carbon and nitrogen in \swap.
\textbf{b} Bond distributions involving carbon in \tmc.
\textbf{c} Bond distributions involving nitrogen in \tmc.
}\label{fig:sup_bond_distributions}
\end{figure}

\begin{figure}[ht]
\centering
\includegraphics[width=1.0\textwidth]{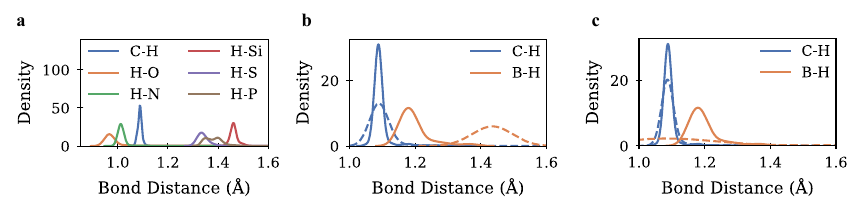}
\caption{
\textbf{Comparison of empirical (solid) and generated (dashed) bond distributions for vanilla and fine-tuned React-OT model.} 
\textbf{a} Empirical bond distributions for a subset of hydrogen containing bonds in \swap.
\textbf{b} Empirical and predicted distribution for C-H and B-H bonds for fine-tuned React-OT model.
\textbf{c} Empirical and predicted distribution for C-H and B-H bonds for vanilla React-OT model.
}\label{fig:sup_bond_distributions_ood}
\end{figure}

Supplementary Figure~\ref{fig:sup_bond_distributions_ood} shows the comparison of the empirical and predicted bonded distribution for the vanilla and \swap fine-tuned React-OT model evaluated on the \ood benchmark. The fine-tuned model overestimates the B–H bond lengths. This behavior can be traced back to the bond-length distributions present in the \swap training data, where bonds between hydrogen and newly introduced elements such as Si, S, and P consistently peak between 1.3 and 1.5~\AA\ (Supplementary Figure~\ref{fig:sup_bond_distributions_ood}a). As a result, the fine-tuned model predicts a broadened B–H distribution centered around 1.45\AA, rather than the empirical peak near 1.2~\AA\ (Supplementary Figure~\ref{fig:sup_bond_distributions_ood}b). By contrast, the vanilla model produces an even broader distribution with a weak maximum near the organic C–H bond length of approximately 1.1\AA\ (Supplementary Figure~\ref{fig:sup_bond_distributions_ood}c), which incidentally leads to a lower overall RMSD. Crucially, this apparent structural agreement of the vanilla model comes at the expense of physical plausibility. Its broader bond-length distributions include a substantial fraction of unphysically short bonds, which is reflected in significantly worse energetic accuracy. Despite achieving lower RMSDs, the vanilla React-OT model exhibits a median energetic error of 27.85~kcal~mol$^{-1}$, compared to 21.54~kcal~mol$^{-1}$ for the \swap-fine-tuned model (Supplementary Table~\ref{tab:sup_performance_ood}). Consistently, evaluation across all bond types confirms inferior distributional agreement for the vanilla model, with a median Wasserstein-1 distance of 0.09\AA, relative to 0.05~\AA\ for the fine-tuned counterpart. A similar trend is observed for \aefm. The \swap fine-tuned model exhibits a modest improvement over the vanilla version, reducing the median RMSD from 0.26~\AA\ to 0.23~\AA, while retaining the same fundamental limitations in extrapolation behavior (Supplementary Table~\ref{tab:sup_performance_ood}).

The performance summary of \aefm on the \swap and \tmc benchmark is provided in Supplementary Table~\ref{tab:sup_performance_aefm}, showing the same qualitative failure modes as React-OT. This is further highlighted in Supplementary Figure~\ref{fig:sup_failure_analysis_aefm}, showing a detailed performance of the vanilla \aefm model on \swap.

\begin{table*}[ht]
\centering
\caption{Structural and energetic errors for variants of \aefm.}
\vspace{0.5em}
\label{tab:sup_performance_aefm}
\begin{adjustbox}{max width=\linewidth}
\begin{tabular}{@{\extracolsep\fill}llllllll}
\toprule
Approach & Dataset & \multicolumn{2}{c}{RMSD (\AA)} & \multicolumn{2}{c}{DMAE (\AA)} & \multicolumn{2}{c}{$|\Delta E_{\text{TS}}|$ (kcal mol$^{-1}$)} \\
\cmidrule(lr){3-4} \cmidrule(lr){5-6} \cmidrule(lr){7-8}
& & Mean & Median & Mean & Median & Mean & Median \\
\midrule
Vanilla AEFM & \swap & 0.52 & 0.47 & 0.28 & 0.25 & 353.68 & 322.44 \\
Vanilla AEFM & \tmc & 1.11 & 1.08 & 0.67 & 0.61 & 2050.31 & 1568.24 \\
Fine-tuned AEFM & \swap & 0.28 & 0.16 & 0.11 & 0.06 & 8.22 & 0.93 \\
Fine-tuned AEFM & \tmc & 0.38 & 0.27 & 0.15 & 0.10 & 9.19 & 1.48 \\
\bottomrule
\end{tabular}
\end{adjustbox}
% \footnotesize
% \justify
% \textsuperscript{\emph{a}} pretraining with 2500 pseudo-reactions of the corresponding dataset. \\
\end{table*}

\begin{figure}[ht]
\centering
\includegraphics[width=0.5\textwidth]
{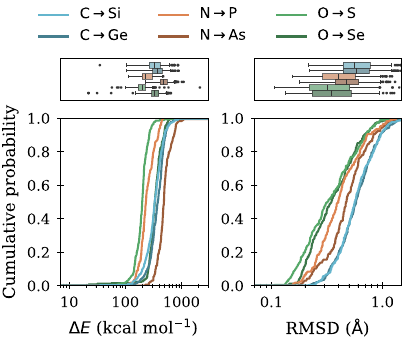}
\caption{
\textbf{Failure modes of the vanilla \aefm model on reactions involving novel chemistry.} 
Cumulative probability distributions of structural and energetic errors for the \swap benchmark, illustrating the sharp performance degradation introduced by single-atom substitutions.
}\label{fig:sup_failure_analysis_aefm}
\end{figure}

\subsection{Data dependency}

\begin{figure}[ht]
\centering
\includegraphics[width=0.5\textwidth]
{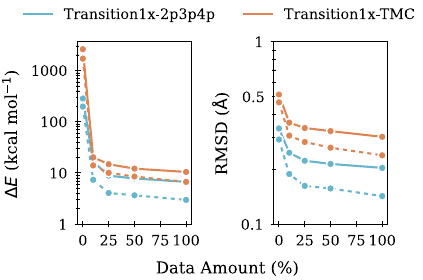}
\caption{
\textbf{Data ablation.} Correlation of energetic error and RMSD for \swap and \tmc as a function of the amount of training data. The case of no additional training data corresponds to the vanilla React-OT model trained on \xtb. Solid lines indicate the mean, while dashed lines indicate the median.
}\label{fig:sup_data_ablation}
\end{figure}

To assess the data dependency of React-OT, we train models on varying fractions of the available altered reaction datasets. Supplementary Figure~\ref{fig:sup_data_ablation} shows the relationship between the energetic error, the structural similarity, and the amount of training data used. The case of zero additional data corresponds to the vanilla React-OT model trained solely on \xtb, whereas increasing percentages indicate the fraction of \swap or \tmc data incorporated during fine-tuning. Solid lines denote the mean error, and dashed lines the corresponding median. A consistent and monotonic reduction in error is observed as the amount of training data increases.

\section{Pseudo-reactions for self-supervised pretraining}

To evaluate whether the generated pseudo-reactions exhibit structural overlap with their corresponding source reactions, that is, the reactions from which the minima were extracted, we compute the structural similarity between each structure in the pseudo-reaction and each structure in the associated source trajectory. The results are reported in Supplementary Tables~\ref{tab:sup_pseudo_reactions_r}–\ref{tab:sup_pseudo_reactions_p}. Notably, even the smallest similarity values remain above 1~\AA, indicating that the pseudo-reactions do not share meaningful structural overlap with real reactions and therefore constitute a distinct pretraining signal.

\begin{figure}[ht]
\centering
\includegraphics[width=1.0\textwidth]
{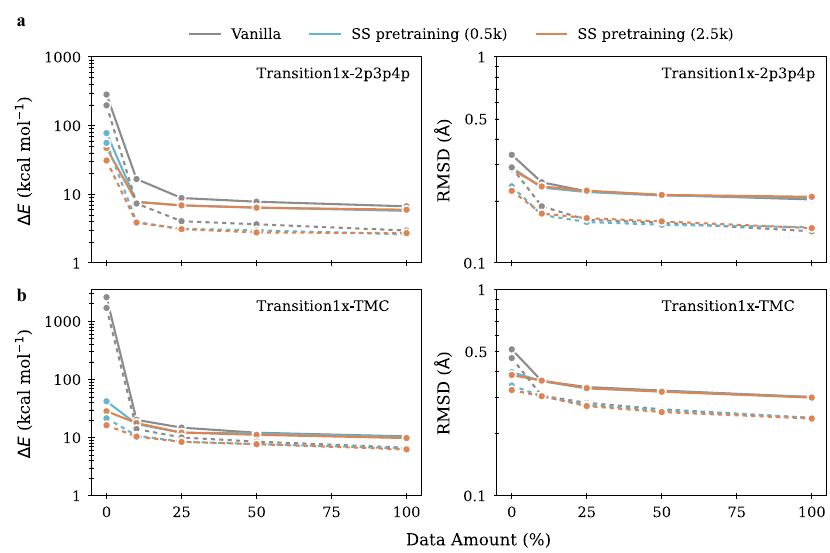}
\caption{
\textbf{Data ablation including self-supervised pseudo-reaction pretraining.} 
\textbf{a}, \textbf{b} Correlation between energetic error and RMSD for \swap and \tmc as a function of training set size. The case without additional training data corresponds to a React-OT model trained on \xtb only, while color denotes whether the model underwent prior self-supervised pretraining using pseudo-reactions constructed from the corresponding dataset. Solid lines indicate mean values, and dashed lines indicate medians.
}\label{fig:sup_crest_data_ablation}
\end{figure}

Supplementary Figure~\ref{fig:sup_crest_data_ablation} presents the relationship between energetic and structural error as a function of the amount of fine-tuning data, comparing the vanilla model to models pretrained on 500 and 2500 pseudo-reactions for (a) \swap and (b) \tmc. For \swap, pretraining on 2500 pseudo-reactions combined with fine-tuning on only 25\% of the data yields performance comparable to a vanilla React-OT model trained on the full dataset. A similar trend is observed for \tmc, although the effect of pretraining is less pronounced than in the \swap case.

\begin{table*}[ht]
\centering
\caption{Similarity of pseudo-reactants to source reactions.}
\vspace{0.5em}
\label{tab:sup_pseudo_reactions_r}
\begin{tabular}{@{\extracolsep\fill}lllllll}
\toprule
Dataset & \multicolumn{2}{c}{RMSD(R$_\text{pseudo}$, R$_\text{ref.}$) (\AA)} & \multicolumn{2}{c}{RMSD(R$_\text{pseudo}$, TS$_\text{ref.}$) (\AA)} & \multicolumn{2}{c}{RMSD(R$_\text{pseudo}$, P$_\text{ref.}$) (\AA)} \\
\cmidrule(lr){2-3} \cmidrule(lr){4-5} \cmidrule(lr){6-7}
& Mean & Median & Mean & Median & Mean & Median \\
\midrule
\swap & 1.92 & 1.37 & 1.89 & 1.33 & 1.24 & 1.28 \\
\tmc & 1.84 & 1.68 & 1.83 & 1.64 & 1.65 & 1.69 \\
\bottomrule
\end{tabular}
\end{table*}

\begin{table*}[ht]
\centering
\caption{Similarity of pseudo-TS to source reactions.}
\vspace{0.5em}
\label{tab:sup_pseudo_reactions_ts}
\begin{tabular}{@{\extracolsep\fill}lllllll}
\toprule
Dataset & \multicolumn{2}{c}{RMSD(TS$_\text{pseudo}$, R$_\text{ref.}$) (\AA)} & \multicolumn{2}{c}{RMSD(TS$_\text{pseudo}$, TS$_\text{ref.}$) (\AA)} & \multicolumn{2}{c}{RMSD(TS$_\text{pseudo}$, P$_\text{ref.}$) (\AA)} \\
\cmidrule(lr){2-3} \cmidrule(lr){4-5} \cmidrule(lr){6-7}
& Mean & Median & Mean & Median & Mean & Median \\
\midrule
\swap & 1.91 & 1.36 & 1.89 & 1.33 & 1.85 & 1.35 \\
\tmc & 1.84 & 1.67 & 1.83 & 1.64 & 1.83 & 1.66  \\
\bottomrule
\end{tabular}
\end{table*}

\begin{table*}[ht]
\centering
\caption{Similarity of pseudo-products to source reactions.}
\vspace{0.5em}
\label{tab:sup_pseudo_reactions_p}
\begin{tabular}{@{\extracolsep\fill}lllllll}
\toprule
Dataset & \multicolumn{2}{c}{RMSD(P$_\text{pseudo}$, R$_\text{ref.}$) (\AA)} & \multicolumn{2}{c}{RMSD(P$_\text{pseudo}$, TS$_\text{ref.}$) (\AA)} & \multicolumn{2}{c}{RMSD(P$_\text{pseudo}$, P$_\text{ref.}$) (\AA)} \\
\cmidrule(lr){2-3} \cmidrule(lr){4-5} \cmidrule(lr){6-7}
& Mean & Median & Mean & Median & Mean & Median \\
\midrule
\swap & 1.18 & 1.24 & 1.23 & 1.23 & 1.24 & 1.28 \\
\tmc & 1.63 & 1.66 & 1.65 & 1.68 & 1.65 & 1.69 \\
\bottomrule
\end{tabular}
\end{table*}

Supplementary Table~\ref{tab:sup_performance_ood} summarizes the impact of pretraining on the ability to elucidate chemically novel reactions. In particular, self-supervised pretraining on 100 pseudo-reactions from the \ood dataset yields clear improvements over the vanilla model trained solely on \xtb data. In contrast, pretraining on \swap alone does not improve performance on these novel systems, underscoring the need for system-specific pretraining strategies. For the catalytic systems, both the vanilla and the pretrained model are trained on 50 target reactions from the respective dataset, while the pretrained model is first trained on 100 pseudo-reactions of the corresponding dataset. Comparing the median energetic errors, shows improvements of 44\% and 33\% for the Pt- and Rh-catalyzed system respectively. A comparison with a React-OT model pretrained on \tmc and subsequently fine-tuned on the same 50 target reactions shows that pseudo-reaction pretraining achieves nearly comparable performance, highlighting its data efficiency.

\begin{table*}[ht]
\centering
\caption{Statistic on additional benchmark systems.}
\vspace{0.5em}
\label{tab:sup_performance_ood}
\begin{tabular}{@{\extracolsep\fill}llllllllll}
\toprule
Approach & Dataset & \multicolumn{2}{c}{RMSD (\AA)} & \multicolumn{2}{c}{DMAE (\AA)} & \multicolumn{2}{c}{$|\Delta E_{\text{TS}}|$ (kcal mol$^{-1}$)} \\
\cmidrule(lr){3-4} \cmidrule(lr){5-6} \cmidrule(lr){7-8}
& & Mean & Median & Mean & Median & Mean & Median & \\
\midrule
Vanilla React-OT & \ood & 0.21 & 0.15 & 0.11 & 0.08 & 41.32 & 27.85 \\
Vanilla React-OT\textsuperscript{\emph{a}} & \ood & 0.20 & 0.14 & 0.09 & 0.06 & 16.87 & 9.22 \\
Vanilla React-OT\textsuperscript{\emph{b}} & \ood & 0.25 & 0.19 & 0.12 & 0.10 & 26.65 & 21.54 \\
Vanilla \aefm & \ood & 0.32 & 0.26 & 0.14 & 0.11 & 26.70 & 16.93 \\
Vanilla AEFM\textsuperscript{\emph{c}} & \ood & 0.30 & 0.23 & 0.13 & 0.08 & 23.28 & 14.08 \\
Vanilla React-OT & Pt-catalyzed & 0.35 & 0.33 & 0.22 & 0.20 & 684.57 & 412.46 \\
Vanilla React-OT\textsuperscript{\emph{a}} & Pt-catalyzed & 0.32 & 0.29 & 0.17 & 0.16 & 185.69 & 130.76 \\
Vanilla React-OT\textsuperscript{\emph{d}} & Pt-catalyzed & 0.34 & 0.30 & 0.18 & 0.17 & 58.38 & 36.52 \\
Vanilla \aefm & Pt-catalyzed & 0.80 & 0.75 & 0.49 & 0.43 & 4928.35 & 931.56 \\
Vanilla AEFM\textsuperscript{\emph{e}} & Pt-catalyzed & 0.31 & 0.25 & 0.17 & 0.13 & 44.05 & 20.92 \\
Fine-tuned React-OT & Pt-catalyzed & 0.20 & 0.16 & 0.09 & 0.08 & 43.19 & 9.04 \\
Fine-tuned React-OT\textsuperscript{\emph{a}} & Pt-catalyzed & 0.19 & 0.15 & 0.08 & 0.07 & 11.42 & 5.06 \\
Fine-tuned React-OT\textsuperscript{\emph{d}} & Pt-catalyzed & 0.20 & 0.15 & 0.09 & 0.07 & 9.67 & 5.68 \\
Vanilla React-OT & Rh-catalyzed & 0.56 & 0.24 & 0.43 & 0.16 & 2065.93 & 374.98 \\
Vanilla React-OT\textsuperscript{\emph{a}} & Rh-catalyzed & 0.20 & 0.19 & 0.18 & 0.17 & 203.16 & 164.66 \\
Vanilla React-OT\textsuperscript{\emph{d}} & Rh-catalyzed & 0.35 & 0.32 & 0.24 & 0.22 & 230.82 & 191.10 \\
Vanilla \aefm & Rh-catalyzed & 0.95 & 0.88 & 0.60 & 0.57 & 3198.51 & 1827.39 \\
Vanilla AEFM\textsuperscript{\emph{e}} & Rh-catalyzed & 0.39 & 0.37 & 0.25 & 0.24 & 172.57 & 150.99 \\
Fine-tuned React-OT & Rh-catalyzed & 0.06 & 0.04 & 0.03 & 0.02 & 38.24 & 3.65 \\
Fine-tuned React-OT\textsuperscript{\emph{a}} & Rh-catalyzed & 0.05 & 0.03 & 0.03 & 0.02 & 26.53 & 2.45 \\
Fine-tuned React-OT\textsuperscript{\emph{d}} & Rh-catalyzed & 0.05 & 0.03 & 0.02 & 0.02 & 7.24 & 1.04 \\
\bottomrule
\end{tabular}
\footnotesize
\justify
\textsuperscript{\emph{a}} OAReactDiff self-supervised pretraining with 100 pseudo-reactions of the corresponding dataset. \\
\textsuperscript{\emph{b}} React-OT pretraining with \swap. \\
\textsuperscript{\emph{c}} \aefm pretraining with \swap. \\
\textsuperscript{\emph{d}} React-OT pretraining with \tmc. \\
\textsuperscript{\emph{e}} \aefm pretraining with \tmc. \\
\end{table*}

\section{Quantum chemical validation}

Supplementary Table~\ref{tab:sup_delta_e_dft} compares the energetic errors obtained when evaluating model predictions with GFN2-xTB and with DFT, where the respective functional and basis set are indicated as superscripts. In all cases, the energetic error is larger at the DFT level, indicating that GFN2-xTB is less sensitive to certain geometric deviations. The \swap and \tmc models are fine-tuned on 100\% of their respective datasets, while the model used for the catalytic sets is pretrained on \tmc and subsequently fine-tuned on 50 target reactions.

Supplementary Table~\ref{tab:sup_ts_opt_dft} reports the structural differences between GFN2-xTB TS geometries and their DFT-optimized counterparts, with energetic differences evaluated consistently at the DFT level. For \swap, GFN2-xTB already provides accurate geometries, requiring only minimal structural adjustments upon DFT optimization (median RMSD of 0.16~\AA). In contrast, both \tmc and the Pt-catalyzed system exhibit larger deviations, reflecting the worse reliability of GFN2-xTB for these reactions.

Supplementary Table~\ref{tab:sup_comparison_ts_dft} compares the React-OT–generated TS structures directly with their corresponding DFT reference TSs, demonstrating that the model achieves reasonably high accuracy, with median energetic errors below 20~kcal~mol$^{-1}$. Finally, Supplementary Table~\ref{tab:sup_comparison_ts_opt_dft} reports the results obtained after DFT optimization of the model-generated structures to valid TSs. Across all evaluated datasets, more than 75\% of generated geometries converge to the correct TS and thereby to the intended reaction pathway, demonstrating the practical utility of the generated structures for downstream quantum-chemical applications.

\begin{table*}[ht]
\centering
\caption{Comparison of energetic error evaluated with GFN2-xTB and DFT.}
\vspace{0.5em}
\label{tab:sup_delta_e_dft}
\begin{tabular}{@{\extracolsep\fill}lllllllll}
\toprule
Approach & Dataset & \multicolumn{2}{c}{$|\Delta E_\text{TS}^\text{GFN2-xTB}|$ (kcal mol$^{-1}$)} & \multicolumn{2}{c}{$|\Delta E_\text{TS}^\text{DFT}|$ (kcal mol$^{-1}$)} \\
\cmidrule(lr){3-4} \cmidrule(lr){5-6}
& & Mean & Median & Mean & Median \\
\midrule
Fine-tuned React-OT & \swap\textsuperscript{\emph{a}} & 6.41 & 2.77 & 8.00 ($\uparrow$25\%) & 3.39 ($\uparrow$22\%) \\
Fine-tuned React-OT & \tmc\textsuperscript{\emph{a}} & 10.37 & 6.81 & 12.78 ($\uparrow$23\%) & 7.79 ($\uparrow$14\%) \\
Fine-tuned React-OT & Pt-catalyzed\textsuperscript{\emph{b}} & 9.67 & 5.68 & 10.87 ($\uparrow$12\%) & 6.89 ($\uparrow$21\%) \\
Fine-tuned React-OT & Rh-catalyzed\textsuperscript{\emph{c}} & 7.24 & 1.04 & 7.75 ($\uparrow$7\%) & 1.21 ($\uparrow$16\%) \\
\bottomrule
\end{tabular}
\footnotesize
\justify
\textsuperscript{\emph{a}} DFT functional/basis set: $\omega$B97x/def2-SVP \\
\textsuperscript{\emph{b}} DFT functional/basis set: PBE0 D3BJ/def2-SVP \\
\textsuperscript{\emph{c}} DFT functional/basis set: $\omega$B97x-D3/def2-SVP \\
\end{table*}

\begin{table*}[ht]
\centering
\caption{Statistics of energetic and structural difference after DFT TS optimization for datasets.}
\vspace{0.5em}
\label{tab:sup_ts_opt_dft}
\begin{tabular}{@{\extracolsep\fill}llllllllll}
\toprule
Dataset & \multicolumn{2}{c}{RMSD (\AA)} & \multicolumn{2}{c}{$|\Delta E_\text{TS}|$ (kcal mol$^{-1}$)} & \# Converged TS & \multicolumn{2}{c}{\# Optimization steps} \\
\cmidrule(lr){2-3} \cmidrule(lr){4-5}
& Mean & Median & Mean & Median & Mean & Median \\
\midrule
\swap\textsuperscript{\emph{a}} & 0.27 & 0.16 & 5.80 & 3.31 & 194/200 & 61 & 43 \\
\tmc\textsuperscript{\emph{a}} & 0.46 & 0.35 & 13.93 & 7.97 & 99/100 & 100 & 80 \\
Pt-catalyzed\textsuperscript{\emph{b}} & 0.42 & 0.33 & 7.19 & 6.19 & 44/50 & 190 & 128 \\
\bottomrule
\end{tabular}
\footnotesize
\justify
\textsuperscript{\emph{a}} DFT functional/basis set: $\omega$B97x/def2-SVP \\
\textsuperscript{\emph{b}} DFT functional/basis set: PBE0 D3BJ/def2-SVP \\
\end{table*}

\begin{table*}[ht]
\centering
\caption{Comparison of energetic and structural difference of React-OT models compared to DFT TS references.}
\vspace{0.5em}
\label{tab:sup_comparison_ts_dft}
\begin{tabular}{@{\extracolsep\fill}lllllll}
\toprule
Approach & Dataset & \multicolumn{2}{c}{RMSD (\AA)} & \multicolumn{2}{c}{$|\Delta E_\text{TS}|$ (kcal mol$^{-1}$)} \\
\cmidrule(lr){3-4} \cmidrule(lr){5-6}
& & Mean & Median & Mean & Median \\
\midrule
Fine-tuned React-OT & \swap\textsuperscript{\emph{a}} & 0.36 & 0.26 & 12.17 & 7.67 \\
Fine-tuned React-OT & \tmc\textsuperscript{\emph{a}} & 0.52 & 0.47 & 21.09 & 16.34 \\
Fine-tuned React-OT & \tmc\textsuperscript{\emph{a},\emph{c}} & 0.52 & 0.42 & 20.11 & 15.48 \\
Fine-tuned React-OT & Pt-catalyzed\textsuperscript{\emph{b}} & 0.49 & 0.46 & 20.48 & 16.99 \\
\bottomrule
\end{tabular}
\footnotesize
\justify
\textsuperscript{\emph{a}} DFT functional/basis set: $\omega$B97x/def2-SVP \\
\textsuperscript{\emph{b}} DFT functional/basis set: PBE0 D3BJ/def2-SVP \\
\textsuperscript{\emph{c}} Self-supervised pretraining with 1500 pseudo-reactions at DFT level of theory. \\
\end{table*}

\begin{table*}[ht]
\centering
\caption{Comparison of DFT refined React-OT samples compared to their DFT TS reference.}
\vspace{0.5em}
\label{tab:sup_comparison_ts_opt_dft}
\begin{adjustbox}{max width=\linewidth}
\begin{tabular}{@{\extracolsep\fill}llllllll}
\toprule
Approach & Dataset & \multicolumn{2}{c}{RMSD (\AA)} & Same (\%) & \# Converged TS & \multicolumn{2}{c}{\# Optimization steps} \\
\cmidrule(lr){3-4} \cmidrule(lr){7-8}
& & Mean & Median & & & Mean & Median \\
\midrule
Fine-tuned React-OT & \swap\textsuperscript{\emph{a}} & 0.20 & 0.00 & 75 & 188/200 & 74 & 51 \\
Fine-tuned React-OT & \tmc\textsuperscript{\emph{a}} & 0.14 & 0.01 & 82 & 93/100 & 96 & 85 \\
Fine-tuned React-OT & Pt-catalyzed\textsuperscript{\emph{b}} & 0.15 & 0.01 & 76 & 37/50 & 162 & 100 \\
\bottomrule
\end{tabular}
\end{adjustbox}
\footnotesize
\justify
\textsuperscript{\emph{a}} DFT functional/basis set: $\omega$B97x/def2-SVP \\
\textsuperscript{\emph{b}} DFT functional/basis set: PBE0 D3BJ/def2-SVP \\
\end{table*}

\FloatBarrier